\RequirePackage{fix-cm}

\documentclass[twocolumn,epjc3]{svjour3}
\smartqed  % flush right qed marks, e.g. at end of proof
\pdfoutput=1 % if your are submitting a pdflatex (i.e. if you have % images in pdf, png or jpg format)

\RequirePackage{graphicx}
\graphicspath{{./figures/}}    % path to figures
\RequirePackage{mathptmx}      % use Times fonts if available on your TeX system
\RequirePackage[numbers,sort&compress]{natbib}
\RequirePackage[colorlinks,citecolor=blue,urlcolor=blue,linkcolor=blue]{hyperref}

\usepackage[usenames,dvipsnames]{xcolor}
\usepackage[normalem]{ulem}   % strikethrough
\usepackage{upgreek}
\usepackage[version=4]{mhchem}  % Chemical element symbols
\usepackage{isotope}
\usepackage{xspace}             % Spaces that don't get eaten
\usepackage{fancyhdr}
\usepackage[switch]{lineno}
\usepackage{url}
\usepackage{float}
\usepackage{caption}
\usepackage{graphicx}
\usepackage{amsmath,amssymb}
\usepackage{nccmath}
\usepackage{lineno}
\usepackage{upgreek}
\usepackage{currfile} % To be able to list current file name in title
\newcommand{\ar}{\isotope[37]{Ar}\xspace}
\newcommand{\kr}{\isotope[83\text{m}]{Kr}\xspace}

\journalname{Eur. Phys. J. C}

\begin{document}

\title{Low-energy Calibration of XENON1T with an Internal $^{\textbf{37}}$Ar Source}



\author{E.~Aprile\thanksref{addr3}
\and
K.~Abe\thanksref{addr23}
\and
F.~Agostini\thanksref{addr0}
\and
S.~Ahmed Maouloud\thanksref{addr18}
\and
M.~Alfonsi\thanksref{addr5}
\and
L.~Althueser\thanksref{addr7}
\and
B.~Andrieu\thanksref{addr18}
\and
E.~Angelino\thanksref{addr14, email1}
\and
J.~R.~Angevaare\thanksref{addr8}
\and
V.~C.~Antochi\thanksref{addr12}
\and
D.~Ant\'on Martin\thanksref{addr1}
\and
F.~Arneodo\thanksref{addr9}
\and
L.~Baudis\thanksref{addr17}
\and
A.~L.~Baxter\thanksref{addr10}
\and
L.~Bellagamba\thanksref{addr0}
\and
R.~Biondi\thanksref{addr4}
\and
A.~Bismark\thanksref{addr17}
\and
A.~Brown\thanksref{addr19}
\and
S.~Bruenner\thanksref{addr8}
\and
G.~Bruno\thanksref{addr13}
\and
R.~Budnik\thanksref{addr16}
\and
T.~K.~Bui\thanksref{addr23}
\and
C.~Cai\thanksref{addr27}
\and
C.~Capelli\thanksref{addr17}
\and
J.~M.~R.~Cardoso\thanksref{addr2}
\and
D.~Cichon\thanksref{addr6}
\and
A.~P.~Colijn\thanksref{addr8}
\and
J.~Conrad\thanksref{addr12}
\and
J.~J.~Cuenca-Garc\'ia\thanksref{addr17,addr26}
\and
J.~P.~Cussonneau\thanksref{addr13,deceased}
% ...
}
\newcommand{\bologna}{Department of Physics and Astronomy, University of Bologna and INFN-Bologna, 40126 Bologna, Italy}
\newcommand{\chicago}{Department of Physics \& Kavli Institute for Cosmological Physics, University of Chicago, Chicago, IL 60637, USA}
\newcommand{\coimbra}{LIBPhys, Department of Physics, University of Coimbra, 3004-516 Coimbra, Portugal}
\newcommand{\columbia}{Physics Department, Columbia University, New York, NY 10027, USA}
\newcommand{\lngs}{INFN-Laboratori Nazionali del Gran Sasso and Gran Sasso Science Institute, 67100 L'Aquila, Italy}
\newcommand{\mainz}{Institut f\"ur Physik \& Exzellenzcluster PRISMA$^{+}$, Johannes Gutenberg-Universit\"at Mainz, 55099 Mainz, Germany}
\newcommand{\heidelberg}{Max-Planck-Institut f\"ur Kernphysik, 69117 Heidelberg, Germany}
\newcommand{\munster}{Institut f\"ur Kernphysik, Westf\"alische Wilhelms-Universit\"at M\"unster, 48149 M\"unster, Germany}
\newcommand{\nikhef}{Nikhef and the University of Amsterdam, Science Park, 1098XG Amsterdam, Netherlands}
\newcommand{\nyuad}{New York University Abu Dhabi - Center for Astro, Particle and Planetary Physics, Abu Dhabi, United Arab Emirates}
\newcommand{\purdue}{Department of Physics and Astronomy, Purdue University, West Lafayette, IN 47907, USA}
\newcommand{\rice}{Department of Physics and Astronomy, Rice University, Houston, TX 77005, USA}
\newcommand{\stockholm}{Oskar Klein Centre, Department of Physics, Stockholm University, AlbaNova, Stockholm SE-10691, Sweden}
\newcommand{\subatech}{SUBATECH, IMT Atlantique, CNRS/IN2P3, Universit\'e de Nantes, Nantes 44307, France}
\newcommand{\torino}{INAF-Astrophysical Observatory of Torino, Department of Physics, University  of  Torino and  INFN-Torino,  10125  Torino,  Italy}
\newcommand{\ucsd}{Department of Physics, University of California San Diego, La Jolla, CA 92093, USA}
\newcommand{\wis}{Department of Particle Physics and Astrophysics, Weizmann Institute of Science, Rehovot 7610001, Israel}
\newcommand{\zurich}{Physik-Institut, University of Z\"urich, 8057  Z\"urich, Switzerland}
\newcommand{\paris}{LPNHE, Sorbonne Universit\'{e}, CNRS/IN2P3, 75005 Paris, France}
\newcommand{\freiburg}{Physikalisches Institut, Universit\"at Freiburg, 79104 Freiburg, Germany}
\newcommand{\napels}{Department of Physics ``Ettore Pancini'', University of Napoli and INFN-Napoli, 80126 Napoli, Italy}
\newcommand{\nagoya}{Kobayashi-Maskawa Institute for the Origin of Particles and the Universe, and Institute for Space-Earth Environmental Research, Nagoya University, Furo-cho, Chikusa-ku, Nagoya, Aichi 464-8602, Japan}
\newcommand{\laquila}{Department of Physics and Chemistry, University of L'Aquila, 67100 L'Aquila, Italy}
\newcommand{\tokyo}{Kamioka Observatory, Institute for Cosmic Ray Research, and Kavli Institute for the Physics and Mathematics of the Universe (WPI), University of Tokyo, Higashi-Mozumi, Kamioka, Hida, Gifu 506-1205, Japan}
\newcommand{\kobe}{Department of Physics, Kobe University, Kobe, Hyogo 657-8501, Japan}
\newcommand{\ucla}{Physics \& Astronomy Department, University of California, Los Angeles, CA 90095, USA}
\newcommand{\kit}{Institute for Astroparticle Physics, Karlsruhe Institute of Technology, 76021 Karlsruhe, Germany}
\newcommand{\tsinghua}{Department of Physics \& Center for High Energy Physics, Tsinghua University, Beijing 100084, China}
\newcommand{\alsoatferrara}{INFN, Sez. di Ferrara and Dip. di Fisica e Scienze della Terra, Universit\`a di Ferrara, via G. Saragat 1, Edificio C, I-44122 Ferrara (FE), Italy}
\newcommand{\alsoatcoimbrapoli}{Coimbra Polytechnic - ISEC, 3030-199 Coimbra, Portugal}
\newcommand{\alsoatuniheidelberg}{Physikalisches Institut, Universit\"at Heidelberg, Heidelberg, Germany}
\newcommand{\mainzreactor}{Forschungsreaktor TRIGA Mainz, Johannes Gutenberg-Universit\"at Mainz, 55099 Mainz, Germany}
%\authorrunning{XENON Collaboration}
\thankstext{deceased}{ Deceased}
\thankstext{addr29}{ Also at \alsoatcoimbrapoli}
\thankstext{addr30}{ Also at \alsoatuniheidelberg}
\thankstext{addr28}{ Also at \alsoatferrara}

% ADD Corresponding Author Email here:
\thankstext{email1}{ e-mail: \texttt{\href{mailto:emanuele.angelino@to.infn.it}{emanuele.angelino@to.infn.it}} (corresponding author) }
\thankstext{email2}{ e-mail: \texttt{\href{mailto:chils@uni-mainz.de}{chils@uni-mainz.de}} (corresponding author) }
\thankstext{email3}{ e-mail: \texttt{\href{mailto:amolinar@to.infn.it}{amolinar@to.infn.it}} (corresponding author) }
\thankstext{email4}{ e-mail: \texttt{\href{mailto:shockley.evan@gmail.com}{shockley.evan@gmail.com}} (corresponding author) }
\thankstext[$\ast$]{emailxe}{ e-mail: \texttt{\href{mailto:xenon@lngs.infn.it}{xenon@lngs.infn.it}}}

\institute{\columbia\label{addr3}
\and
\tokyo\label{addr23}
\and
\bologna\label{addr0}
\and
\paris\label{addr18}
\and
\mainz\label{addr5}
\and
\munster\label{addr7}
\and
\torino\label{addr14}
\and
\nikhef\label{addr8}
\and
\stockholm\label{addr12}
\and
\chicago\label{addr1}
\and
\nyuad\label{addr9}
\and
\zurich\label{addr17}
\and
\purdue\label{addr10}
\and
\lngs\label{addr4}
\and
\freiburg\label{addr19}
\and
\subatech\label{addr13}
\and
\wis\label{addr16}
\and
\tsinghua\label{addr27}
\and
\coimbra\label{addr2}
\and
\heidelberg\label{addr6}
\and
\kit\label{addr26}
\and
\laquila\label{addr22}
\and
\rice\label{addr11}
\and
\napels\label{addr20}
\and
\nagoya\label{addr21}
\and
\ucsd\label{addr15}
\and
\kobe\label{addr24}
\and
\mainzreactor\label{addrXX}
}

\date{Received: date / Accepted: date}

\maketitle

\begin{abstract}
\sloppy
A low-energy electronic recoil calibration of XENON1T, a dual-phase xenon time projection chamber, with an internal \ar source was performed. This calibration source features a 35-day half-life and provides two mono-energetic lines at 2.82 keV and 0.27 keV. The photon yield and electron yield at 2.82 keV are measured to be ($32.3\,\pm\,0.3$) photons/keV and ($40.6\,\pm\,0.5$) electrons/keV, respectively, in agreement with other measurements and with NEST predictions. The electron yield at 0.27 keV is also measured and it is ($68.0^{+6.3}_{-3.7}$) electrons/keV.  The \ar calibration confirms that the detector is well-understood in the energy region close to the detection threshold, with the 2.82 keV line reconstructed at ($2.83\,\pm\,0.02$) keV, which further validates the model used to interpret the low-energy electronic recoil excess previously reported by XENON1T. The ability to efficiently remove argon  with cryogenic distillation after the calibration proves that \ar can be considered as a regular calibration source for multi-tonne xenon detectors. 
\end{abstract}

\section{Introduction}
\label{sec:intro}
\sloppy
The XENON1T experiment, located underground at the INFN Laboratori Nazionali del Gran Sasso (LNGS) in Italy, was designed for direct detection of dark matter (DM) in the form of weakly interacting massive particles (WIMPs)\,\cite{RefXE1TDMExp,RefXE1TDMResSR1}. The detector was a liquid xenon dual-phase time projection chamber (LXe TPC), hosting a liquid volume with a small layer of gaseous xenon on top. It operated with a total of 3.2 t of ultra-pure xenon. It was surrounded by a stainless steel tank filled with water which operated as an active Cherenkov muon veto against cosmic radiation \cite{RefMuonVeto}.
The cylindrical LXe TPC, with a height of 97 cm and a diameter of 96 cm, contained 2\,t of LXe as a target. It was instrumented with 248 photomultiplier tubes (PMTs) arranged in two arrays at the top and bottom to efficiently detect the signals \cite{RefPMTs}.
An electric drift field was applied between a cathode at the bottom of the LXe TPC and a grounded electrode at the top (named "gate"). With a third electrode, the  anode located in the gaseous phase, an extraction field was applied.

In LXe, the total energy deposited by interacting particles is divided into excitation, ionization and heat \cite{RefHitachi,RefAprile2005,RefManzur,RefSorensen,RefDahlThesis}. %The heat can not be detected with LXe TPCs and the fraction of energy in case of a \ar interaction lost in this channel can be neglected. (explained later) 
The detectable energy is split between excitation and ionization, causing these channels to be anti-correlated, whereby the ratio of energy splitting is defined by the applied drift field. Some of the freed electrons from ionization recombine with Xe ions. Excited Xe atoms and Xe ions formed from recombination create Xe dimers which emit UV scintillation light (S1). The remaining free electrons are drifted to the gate by the applied field and extracted into the gas phase, generating a secondary scintillation signal (S2) that is proportional to the number of extracted electrons.

The light distribution pattern of the S2s on the top PMT array provides $x$-$y$ coordinates ($x,y=0$ at the center of the TPC), while the time difference between the S1 and S2 signals is used to measure the depth of the interaction $z$ ($z=0$ at the gate). This allows for a full 3D position reconstruction of the events. Furthermore, the ratio between the S2 and S1 signals makes it possible to discriminate between electronic recoils (ERs) and nuclear recoils (NRs). ERs are induced by $\upbeta$-particles, $\upgamma$-rays or neutrinos scattering off electrons, while NRs are induced by neutrons or neutrinos scattering off nuclei. WIMPs are expected to induce NRs, while other DM candidates can generate ERs. The difference in the S2/S1 ratio between ERs and NRs is caused by the different ionization densities at the interaction sites deposited by the two interaction types, thus resulting in different excitation-to-ionization and recombination ratios in the LXe. %In addition, for ERs, the fraction of energy lost to heat, which is not detectable in a LXe TPC, is independent of energy and recombination, and it does not impact the event reconstruction \cite{RefDahlThesis}.
For ERs, the empirically-determined average energy required to produce a quantum (scintillation photon or ionization electron) is independent from the energy scale and includes the fraction of energy dissipated into heat, which is not detectable in LXe. For NRs, a quenching factor is considered to account for the larger fraction of energy lost into heat compared to the electronic channel \cite{RefLenardo2015}.

In XENON1T, the unprecedentedly low background level of $(76 \pm 2)\,\text{events}\,/(\text{t}\,\cdot\,\,\text{y}\,\cdot\,\text{keV})$ for ERs between 1\,keV and 30\,keV opened up the possibility of investigating alternative DM candidates and other physics beyond the Standard Model. A significant excess with respect to background expectations was found in the low-energy ER spectrum between 1\,keV and 7\,keV\,\cite{RefXE1TExcessER}. The shape of the excess statistically allowed several possible explanations: solar axions, bosonic dark matter or neutrinos with enhanced magnetic moment. A trace amount of tritium is also able to explain the excess, % however its presence in the TPC could neither be confirmed nor excluded. 
the data however is also compatible with a null hypothesis. This is also valid for a possible contamination with \ar, but its concentration could be tightly constrained and it was found to be negligible.

In XENONnT, the latest phase of the XENON project, no excess above background has been found, excluding interpretations of the XENON1T excess as due to physics beyond the Standard Model \cite{RefXEnTER}. Despite this recent result, a further investigation of the XENON1T excess is still valuable, and it requires a robust description of the behaviour of the detector and validation of the energy reconstruction in the region of interest.

It was important to calibrate the low-energy ER region with high statistics in the whole TPC volume. Due to the size of the LXe TPC and the self-shielding properties of LXe, external calibration sources were unable to probe the internal regions of the detector. In XENON1T two internal calibration sources were primarily used. \kr, a source with a two-step decay producing lines at 32 keV and 9 keV, was used to study the spatial response of the detector\,\cite{RefXE1TAnalysis1}. The $\upbeta$-decay of $^{212}$Pb (via injection of $^{220}$Rn) was used to characterize the low-energy ER region, but produces a continuous spectrum, which complicates studies of spatial dependence\,\cite{RefLangRn}. Therefore, a low-energy, uniformly-distributed, and mono-energetic calibration source was desired to validate the detector response in the low energy region where the XENON1T excess was observed. 

%\ar decays with a half-life of 35.01 days via electron capture into $^{37}$Cl under emission of an Auger-Meitner electron, accompanied by a cascade of x-rays~\cite{RefTable37Ar}. Its decay provides calibration lines at energies of 2.82 keV for K-shell (90.21\% probability) and 0.27 keV for L-shell transitions (8.72\% probability) \cite{RefArtificialAr37NeutrinoSource}, allowing for studies of the detector response at these low energies\,\cite{RefPixey}. %The advantage over $^{220}$Rn is that two mono-energetic peaks are available instead of a continuous band.

\ar decays by electron capture into $^{37}$Cl (Q-value of 813.9 keV) with half-life of 35.01 days, emitting a mono-energetic neutrino \cite{RefTable37Ar}. The vacancy, which follows the electron capture from one of the shells (K-, L- or M-shell) is filled by an electron rearrangement accompanied by a cascade emission of x-rays and Auger-Meitner electrons. %  It follows, in the shell where electron was captured (K-, L- or M-shell), a vacancy  whose filling produces an electron rearrangement accompanied by a cascade emission of x-rays and Auger-Meitner electrons. 
The overall energy deposits, corresponding to the K-, L- and M-shell electron binding energies, result in lines at 2.82 keV, 0.27 keV and 0.01 keV, with a branching ratio of 90.2\%, 8.7\% and 1.1\%, respectively \cite{RefAr37energy}.

The \ar K- and L-shell  lines allow to study the low-energy detector response \cite{RefPixey,RefXurich37Ar} , since they are well below the low-energy peak of \kr at 9 keV and close to the 1 keV energy threshold of XENON1T. In general, K-shell decays produce detectable S1s and S2s, while at L-shell energy only S2s are expected. For this reason, two different analyses are required and performed in the following. \iffalse In the S1-S2 analysis only the K-shell events are used, as no direct scin4141tillation light (S1) is expected for the low energetic L-shell interactions. The L-shell events are studied in the S2-only analysis, which also includes K-shell events for verification of the results. \fi

This paper describes the calibration with \ar performed for the first time in XENON1T at the end of the last science run and it is organized as follows: the \ar source production, injection in the detector and removal after the end of the calibration period are described in section \ref{sec:2}. In section \ref{sec:s1s2} the standard analysis of \ar K-shell signals,  making use of both S1 and S2, is presented. The impact of these results on the XENON1T low-energy ER excess is discussed in section \ref{sec:enrec}. Section \ref{sec:s2only} focuses on the S2-only analysis which utilises only the ionization signal, lowering the energy threshold down to the energy of the L-shell decays at 0.27 keV. Conclusions are reported in section \ref{sec:conclusions}.

\section{Source production, injection and removal}
\label{sec:2}
\subsection{Source production}
The isotope \ar is obtained by irradiation of argon enriched in $^{36}$Ar to 99.99 \%, contained in a quartz glass container (ampule), with thermal neutrons:
\begin{linenomath*}
\begin{ceqn}
\begin{equation}
^{36}\text{Ar}(n,\gamma)^{37}\text{Ar}.
\end{equation}
\end{ceqn}
\end{linenomath*}
The enrichment enhances the production of \ar and reduces the production of undesired isotopes. The sample was irradiated at the TRIGA Mark II reactor located at the University of Mainz, which provides a flux of thermal neutrons of $4.2 \times 10^{12}\, \text{n}\,/\,(\text{cm}^{2}\,\cdot\,\text{s})$ \footnote{https://www.kernchemie.uni-mainz.de/reaktor/technische-daten/}. The total activity produced during the irradiation was (100 $\pm$ 10) kBq. This value was calculated based on the neutron flux, the neutron capture cross-section of $^{36}$Ar and the irradiation time. The activity produced cannot be directly measured as the low-energy electrons and x-rays do not penetrate the walls of the quartz-ampule. 

\subsection{Dosing and insertion into XENON1T}
\sloppy
A rate of $O$(10) Hz is needed in the detector to achieve a reasonable amount of data, much lower than the overall activity produced during irradiation. A dosing system is used to open the ampule with a guillotine mechanism. It is also used to reduce the activity to a suitable calibration dosage and inject the irradiated argon into the LXe TPC via the XENON1T purification system. Besides the device to open the ampule and store the \ar, the dosing system consists of an arrangement of sections of well-defined volumes that are used to successively dilute a small portion of the activity with xenon. The amount of \ar injected can be computed from the known volume sizes and the starting activity.
With an initial activity of about 100 kBq, the dosing systems design enables doses of $O$(Bq) to be injected into the detector.

\begin{figure}[h!]
  \centering
  \includegraphics[width=0.5\textwidth]{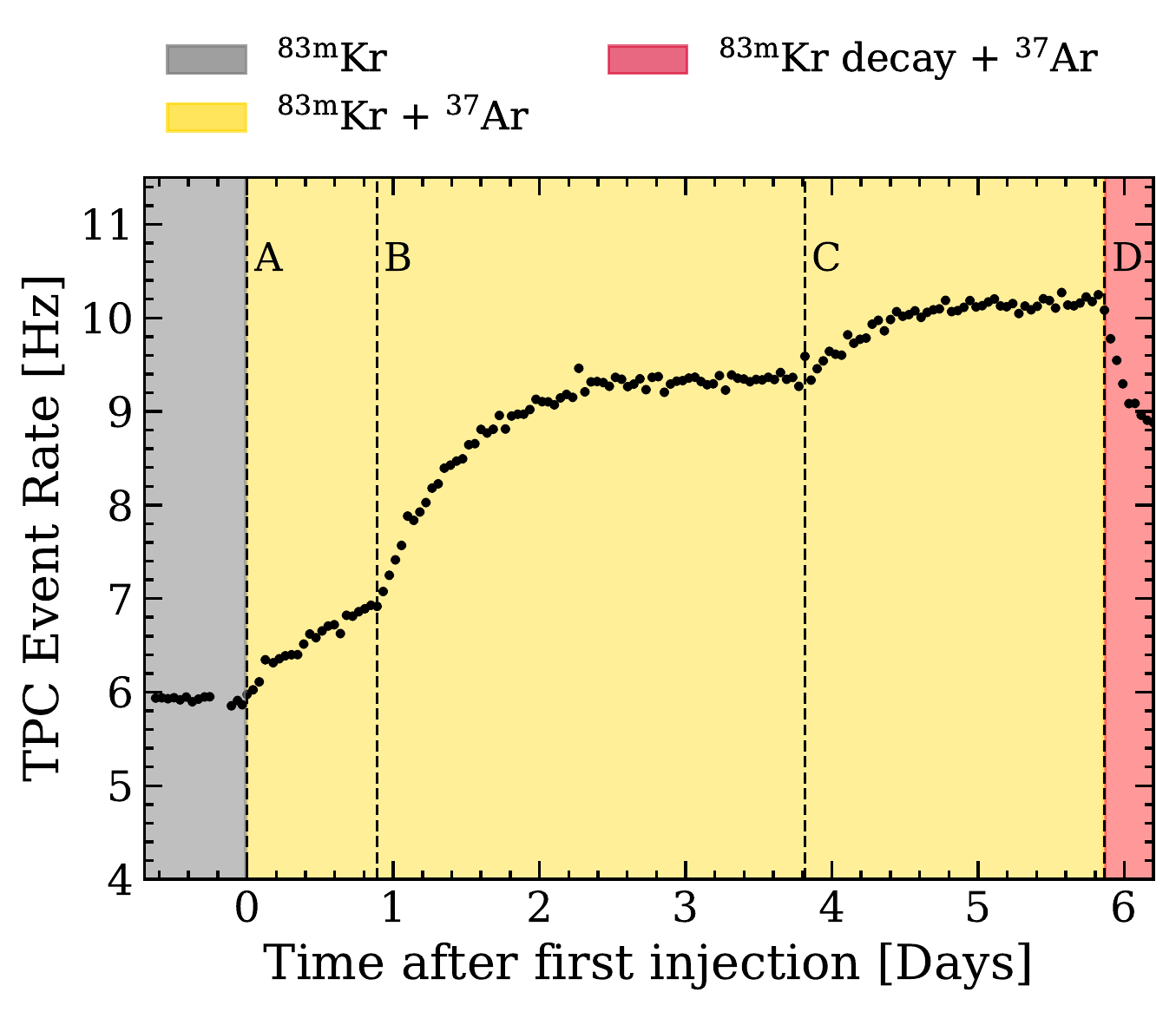}
  \caption{Injection phase of the \ar calibration with the activity increase visible after each of the three injections with expected rates of 2.2, 8.0 and 3.9 Bq (Timestamps A, B and C, respectively), with estimated uncertainties of 10\%. The \ar injections were added on top of a continuous \kr injection with a constant rate of 6 Bq. %Prior to the first \ar injection, \kr was already injected (gray region). 
  This allowed recording of \ar and \kr simultaneously for direct comparison of both sources at identical detector conditions (yellow region). Each data point shows the time-averaged rate during a one-hour long data set of all events detected in the complete TPC. The rate drop at the end (red region) is caused by the decay of the \kr, after the continuous supply of this source was stopped (D).}
  \label{fig:ArInj}   
\end{figure}

The evolution of the detector trigger rate is shown in Fig.\,\ref{fig:ArInj}, without any data selection applied. The flat rate in the beginning is due to the \kr source which was injected in parallel to allow simultaneous calibration with both sources. 
This simultaneous calibration is possible due to the difference in decay energy of both sources, which allows a clean separation of the respective events.
The \ar calibration started with a computed dose of 2.2 Bq injected on October 22nd, 2018. The injection was immediately visible as an increase in the event rate of the detector and is marked as day zero in Fig. \ref{fig:ArInj}. Two additional injections aimed at 8.0 Bq and 3.9 Bq, respectively, were performed in the following days to ensure sufficient statistics for the later analysis. The uncertainties on these values are estimated to be around 10\%, dominated by the initial uncertainty in the activity produced during ampule irradiation.
Inconsistencies between the expected doses and the rate increase shown in Fig. \ref{fig:ArInj} can be caused by unknown systematics from the dosing system and the fact that different dosing procedures were tested.
Injecting small activities at first avoids an accidental overdose and allows verification that the produced activity is in the expected order of magnitude. However a precise knowledge of the absolute activity is not necessary to perform the detector calibration. 

After data processing, an energy cut in the S2 vs S1 parameter space was applied to select the 2.82\,keV \ar peak, together with a fiducial volume cut to reduce background events. More details on the selection can be found in Sec. \ref{sec:s1s2DataSel}. A detailed model~\cite{RefXE1TOnlDist} was fitted to the data and yielded injected activities of (2.6 $\pm$ 0.1) Bq, (4.3 $\pm$ 0.1) Bq and (2.1 $\pm$ 0.1) Bq. Note that this model estimates the activity injected into the purification system, while Fig.\,\ref{fig:ArInj} only shows the rate increase caused by the injection of argon inside the sensitive, liquid filled volume of XENON1T. %At this point as expected  \ar was uniformly distributed in the LXe TPC.% as a result of the xenon recirculation.

\subsection{$^{37}$Ar removal by cryogenic distillation}

\begin{figure*}[h!]
  \centering
  \includegraphics[width=0.95\textwidth]{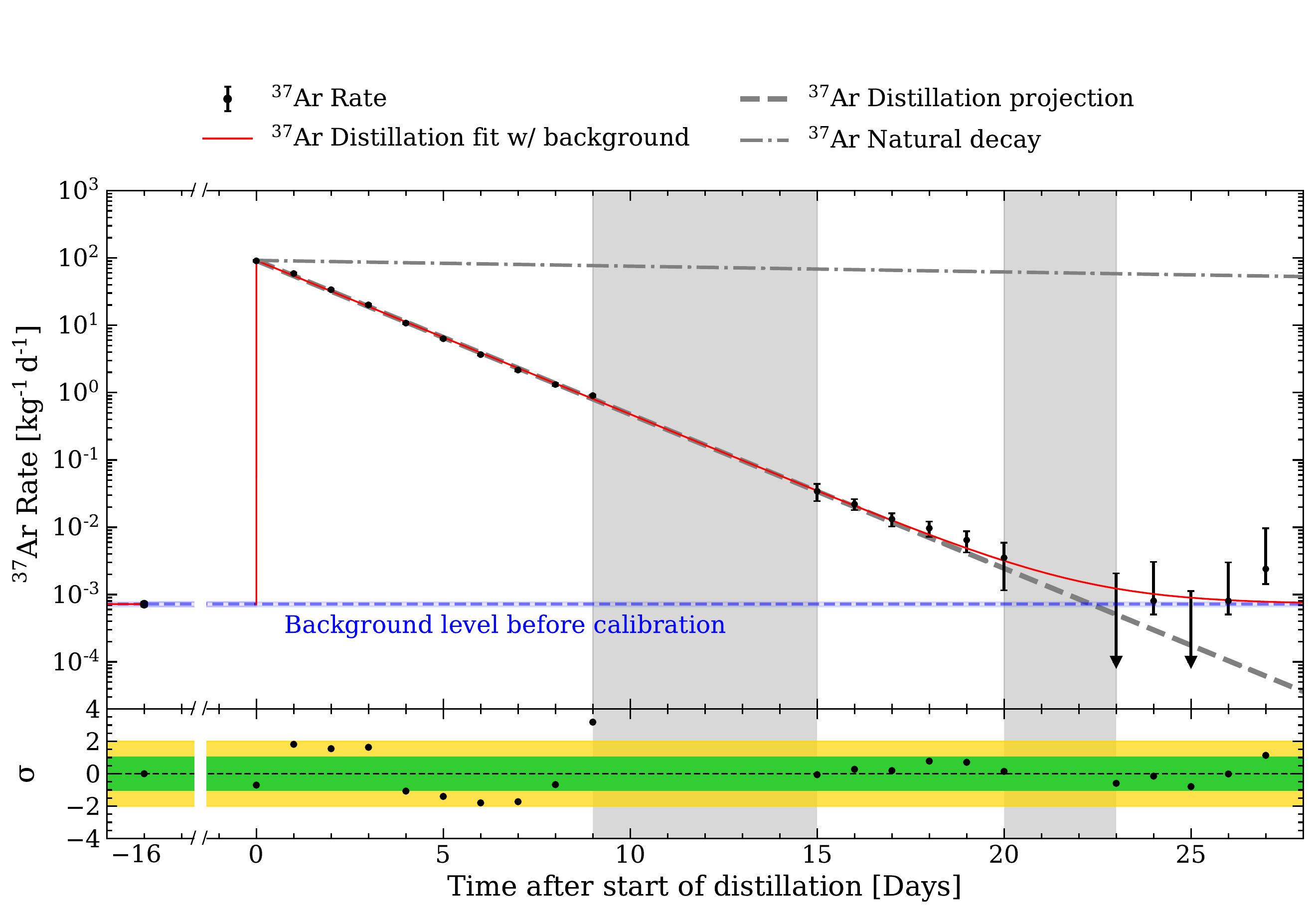}
  \caption{Time evolution of the \ar K-shell event rate in one-day averages during the distillation campaign which started 16 days after the first injection. The red line presents a binned likelihood fit of an exponential plus flat background. The background level is measured over the course of the entire previous science run before the first injection, with correspondingly small error, and is shown here at day -16 and with the dashed blue line. The dashed gray line shows a projection of the \ar rate without taking background into account. The downward arrows represent days with no \ar events detected. The regions in the gray bands are excluded to avoid contamination with background events caused by other calibrations at that time. For comparison, the expected \ar rate development based on natural decay alone is also shown with a dashed-dotted gray line. 
  The uncertainties for the low rates are computed using Feldman-Cousins intervals \cite{RefFeldCousInt}. The lower panel shows the standardized residuals, computed by dividing the difference between data and model by the $\upsigma$ of the data. The green and yellow bands show the 1 and 2 $\upsigma$ deviation, respectively.
  \label{fig:ArDist}
  }
\end{figure*}

The \ar half-life time of 35.01 days makes active removal of the isotope after a calibration run necessary, since a reduction of at least four orders of magnitude is required to return to normal background level. Due to the higher volatility of argon compared to xenon, it accumulates in the gaseous phase and makes removal via cryogenic distillation effective. Removal was performed using the distillation column initially designed to remove $^{85}$Kr, one of the major background sources of XENON1T~\cite{RefXE1TOnlDist,RefXE1TKrCryoDest}. Distillation was conducted online during normal detector operation which allowed for continuous removal of \ar from the gaseous phase. 

The time evolution of the \ar K-shell event rate during distillation is shown in Fig. \ref{fig:ArDist}.
With the same selection criteria and cuts used in the injection phase, a simple exponential function, accounting for both \ar decay and removal by distillation, was used to fit the process. The removal time constant found for the distillation is $\tau_{Ar, dist} = (1.97\,\pm\,0.10)$ days. This means a reduction of the \ar activity by an order of magnitude every (4.54$\,\pm\,$0.23) days. For this case, background level conditions were reached again after 24 days of distillation when the \ar activity has been reduced by five orders of magnitude and can no longer be distinguished from the background level measured before the first injection. For comparison, a waiting time of approximately 570 days would have been necessary to reach the same low \ar activity level via natural decay. The last data point in Fig. \ref{fig:ArDist} exhibits a higher uncertainty because it refers to a day in which the livetime was lower. This last measurement is at 1.1$\upsigma$ from the expected background level. However, there are no data available after this point to track the trend of the rate on longer time scales. 

The difference for the time constant to the value of $\tau_{Ar} = 1.7$ days obtained in Ref.~\cite{RefXE1TOnlDist} is caused by the extended dataset used, which is taking two more weeks of distillation into account. Additional testing of the detector was performed before the start of distillation. Due to this the detector was transitioning into a new equilibrium. The resulting instabilities, which cannot be modeled with this simple exponential function, caused the large residuals shown in Ref.~\cite{RefXE1TOnlDist}.

 Since it was possible to remove the source in a reasonable timescale, \ar should be considered as a regular calibration source for multi-tonne xenon detectors equipped with on-site distillation capabilities.

\section{S1-S2 analysis}
\label{sec:s1s2}

In this analysis, \ar K-shell events are considered where both signals, S1 and S2, are observed. The goal is to determine the absolute photon yield (PY, ph/keV) and electron yield (EY, e$^-$/keV) of the 2.82\,keV energy deposition as well as the linearity of the detector response down to $O$(keV) energies.

As mentioned in Sec. \ref{sec:intro} the energy deposited by an ER is shared between the scintillation and ionization channels as
\begin{linenomath*}
\begin{ceqn}
\begin{equation}
    E = W ( n_{\gamma} + n_e ),
    \label{eq:quanta_basic}
\end{equation}
\end{ceqn}
\end{linenomath*}
where $W$ is the average energy required to produce one quantum (photon or electron), while $n_{\gamma}$ and $n_{e}$ are the number of produced photons and electrons, respectively. For consistency with other XENON data analyses, $W=13.7\pm0.2~\rm eV/quanta$ \cite{RefDahlThesis} is used here, although other recent works suggest smaller values of $W=11.5^{+0.2}_{-0.3}~\rm eV/quanta$ \cite{RefExcEnBaudis, RefExoExcEnAnton}. The choice of a different $W$ value impacts the determination of $n_{\gamma}$ and $n_e$ and the values of detector dependent constants $g_1$ and $g_2$, which parametrize the average size of the S1 and S2 signals per photon or electron, respectively, but it does not affect the energy reconstruction itself.
The S1 (S2) signals in the XENON1T LXe TPC are proportional to the number of photons (electrons) produced at the interaction site. Therefore, Eq. \eqref{eq:quanta_basic} can be converted to observable quantities:
\begin{linenomath*}
\begin{ceqn}
\begin{equation}
    \hat{E} = W \left( \frac{\mathrm{cS1}}{g_1} +\frac{\mathrm{cS2}}{g_2} \right),
    \label{eq:quanta_observable}
\end{equation}
\end{ceqn}
\end{linenomath*}
where \iffalse two detector-specific gain constants $g_1$ and $g_2$ parametrize the average size of the S1 and S2 signals per photon or electron, respectively, and \fi $\hat{E}$ denotes reconstructed energy. The S1 and S2 signals are corrected for position-dependent effects including light collection efficiency (for both S1 and S2). Additionally S2 is corrected for the loss of electrons due to attachment to electronegative impurities, which is a function of the drift time and follows an exponential law, with the electron lifetime as a decay parameter \cite{RefXE1TAnalysis1}. These corrections lead to the cS1 and cS2 variables seen in Eq. \eqref{eq:quanta_observable}.

The $g_1$ and $g_2$ values, assuming a linear detector response, relate the detected light (S1) and charge (S2) yields to the intrinsic PY and EY for a given drift field:

%Once the $g_1$ and $g_2$ values are known -- and assuming a linear detector response -- it is straightforward to convert from an average detected S1 (S2) signal to the average produced number of photons (electrons) by simply scaling the distribution mean by the respective gain constant. That is, for a given electric field, the intrinsic photon (electron) yield PY (EY) is given by:
\begin{linenomath*}
\begin{ceqn}
\begin{align}
\mathrm{PY}(E) &= L_y(E) / g_1, \\
\mathrm{EY}(E) &= Q_y(E) / g_2,
\end{align}
\end{ceqn}
\end{linenomath*}
where $L_y$ ($Q_y$) is the light (charge) yield, i.e. the average cS1 (cS2) per unit energy $E$. The goal of this analysis is to first determine $L_y$ and $Q_y$ of the \ar K-shell, and then extract $g_1$ and $g_2$ to calculate the PY and EY of LXe at the XENON1T drift field of $82$ V/cm \cite{RefXE1TAnalysis}. The value of $g_2$ depends on how S2 is chosen. In the S1-S2 analysis in this paper the corrected S2 signal from the bottom PMT-array only is used. 

\subsection{Data selection}
\label{sec:s1s2DataSel}

The data used in this analysis is from a period including roughly 3.2 days of the injection phase as shown in Fig.~\ref{fig:ArInj}. %During this time, the calibration source \kr was also continuously injected into the detector. \kr is uniformly distributed throughout the active volume and produces mono-energetic depositions that are easy to select in analysis. Calibrating with both sources simultaneously allows for cross-checks of the analysis methods.

The calibration sources can be distinguished in the S1-S2 parameter space. For \ar, all events with $0\,< \text{S1} < 40~\text{photoelectrons (PE)}$ and $500\,< \text{S2} < 2500~\text{PE}$ are selected. A fiducial volume cut of $-93\,<\,z\,<\,-9\,$ cm and radius $r < 42$ cm is also applied for the main results of this analysis, as described in Ref.~\cite{RefXE1TExcessER}.

The standard pre-selections of data are applied in order to exclude events happening when the data acquisition was in a busy condition, which may result in data loss, or when a single PMT detects a large fraction of the total signal. Further data quality cuts are applied. These include selections to ensure correct pairing of S1-S2 signals and to remove misclassified events caused by single electron emission, events dominated by one PMT, or events with an unreliable $x$-$y$ position reconstruction. %Additionally, the signal shapes are taken into account to remove anomalous events or events with a PMT hit-pattern deviating from expectations. 
A more complete list of all applied selections with detailed explanations can be found in Ref.~\cite{RefXE1TAnalysis1}.

\subsection{Data correction}
\label{sec:datacorr}
Several corrections are applied to the data to account for non-uniform detector response throughout the LXe TPC ~\cite{RefXE1TAnalysis1}. The S1 signal is corrected to account for position-dependent light collection efficiency, particularly as a function of depth ($z$) but also of $x$ and $y$. The S2 signal is corrected for both (a) charge loss as a function of $z$ due to electronegative impurities and (b) light collection efficiency variation as a function of $x$ and $y$, due to geometric effects on the S2 gain.% (e.g. from electrode sagging). 

The corrections described above are determined from \kr calibration and applied in all XENON1T analyses. Using the dataset collected with both \ar and \kr injected simultaneously, the same corrections are extracted using \ar and compared to the standard results from \kr, based on earlier calibrations. The electron lifetime extracted from \ar is approximately 10\% higher than the one from \kr. This discrepancy, which was also observed in Ref.~\cite{RefXE1TAnalysis1}, is attributed to a slight non-uniformity of the drift field in the LXe TPC. Such an effect leads to position-dependent variations in the light and charge yields.

Indeed, the non-uniform drift field produces variations in the number of produced electrons and, consequently, the detected S2. This effect is not separable from the variation of S2 due to electrons absorption by impurities.
Due to the different impact of drift field at different energies on charge yield variations, the estimation of electron lifetime for various sources may differ. Hence, in order to evaluate the true electron lifetime, energy-dependent correction should be applied for specific calibration sources.

%In particular, the way in which the charge yield changes, due to the variations of the drift field, depends on the energy deposited. 
%This implies that the variations of the produced S2 across the TPC are different for different sources, resulting in different estimated electron lifetimes. This means that the correction for this specific effect depends on the energy of the specific calibration source.
%The yields depend on the field, on the energy and on the interaction type of the event, resulting in source-dependent corrections. 

For the S1-S2 analysis described in this section, the electron lifetime based on \kr is used in order to be consistent with other calibration points in this science run (see Fig.~\ref{fig:dokeFinal}) as well as other XENON1T analyses. While a different choice of the electron lifetime impacts on the cS2 and $g_2$ values, the energy reconstructed with Eq. \eqref{eq:quanta_observable} is not affected.

\begin{figure}[h!]
  \centering
  \includegraphics[width=0.5\textwidth]{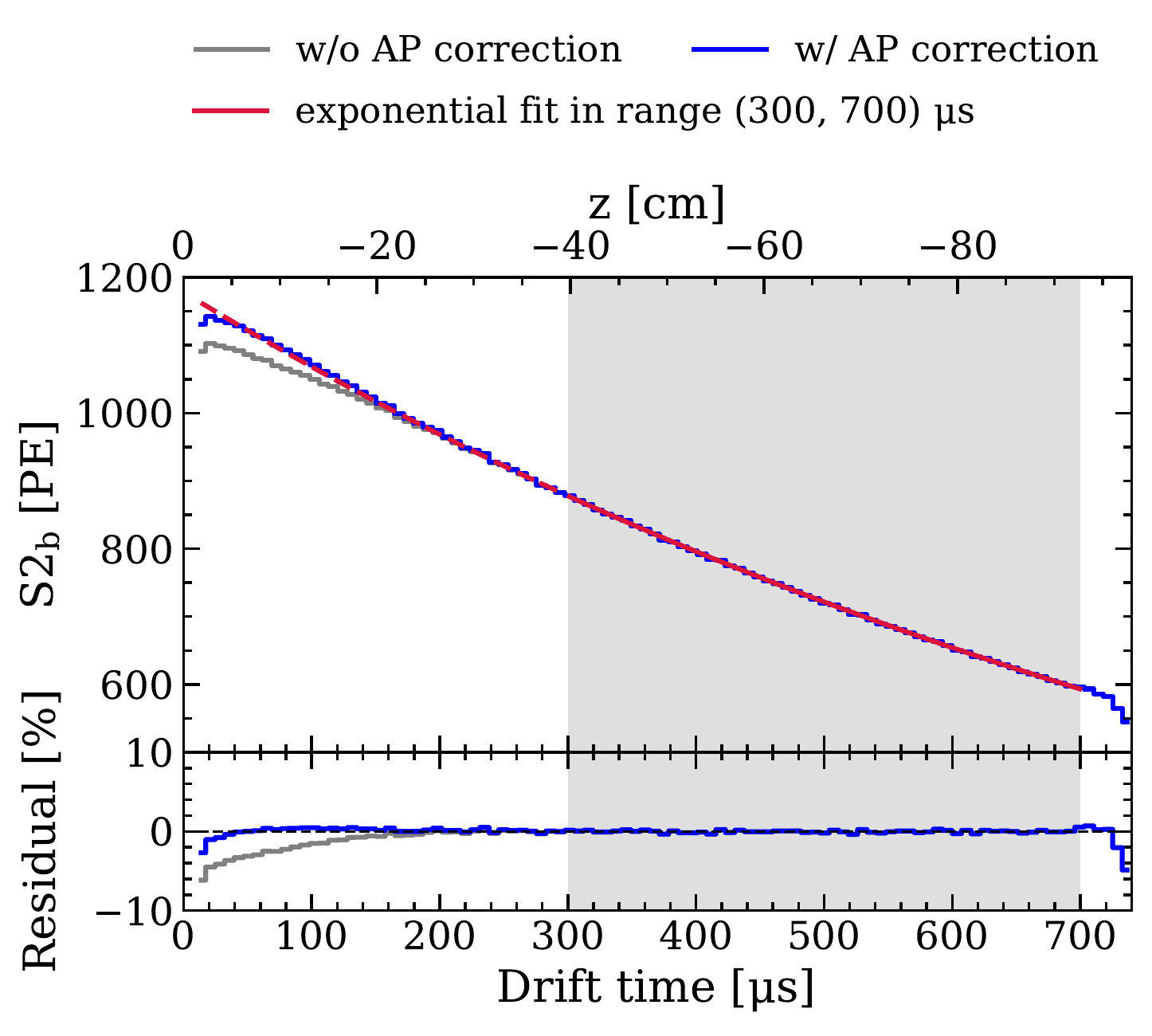}
  \caption{Electron lifetime estimation from \ar data. Without applying an additional correction, the expected exponential decrease with drift time does not well describe the S2 sizes near the top of the detector (low drift times). The gray line shows the S2$_b$ (size of the S2 signal recorded in the bottom PMT array) while the blue one shows the S2$_b$ after applying the afterpulses
  correction. The red line is an exponential fit to the afterpulse corected data to extract the electron lifetime. The region chosen for the exponential fit is denoted in gray.}
  \label{fig:elife}   
\end{figure}

During the electron lifetime analysis, another effect was observed that motivated an additional correction for \ar data: in the top $\sim$20 cm of the LXe TPC, the size of the observed S2 signals were on average a few percent smaller than expected, as shown in Fig.~\ref{fig:elife}. The source of this bias was determined to be PMT afterpulsing. 
Afterpulses are due to residual gas molecules in the PMT vacuum that get ionized by a photoelectron and then drift to the photocathode generating a delayed signal. Afterpulses occur within a few microseconds after the primary S2 peak and thus are typically merged together with it~\cite{RefPmtCharact}. 
As the afterpulse size is directly proportional to the primary S2 size, this results in a small but constant bias that is absorbed into the value of $g_2$, which also applies to all comparable signals.  Due to the low energy of the \ar interaction the energy is deposited in a small volume, resulting in short S2. Additionally, at the top of the LXe TPC, the broadening by diffusion is reduced, so many \ar S2 signals occuring there were reconstructed separately from their afterpulses, resulting in a relatively smaller S2 value. 
An additional correction is made on these events that have secondary S2s within 5\,$\mu$s of the primary and with total area larger than 60 PE ($\sim$2 extracted electrons). For events with this topology, the areas of the primary and secondary  S2 signals are summed up to a total S2 including the afterpulse area. This afterpulse correction helps to ensure a linear energy scale across all energies and across the full fiducial volume. The effect of this correction is shown in  Fig. \ref{fig:elife}.

\subsection{Data analysis and results}

\begin{figure}[h!]
  \centering
  \includegraphics[width=0.5\textwidth]{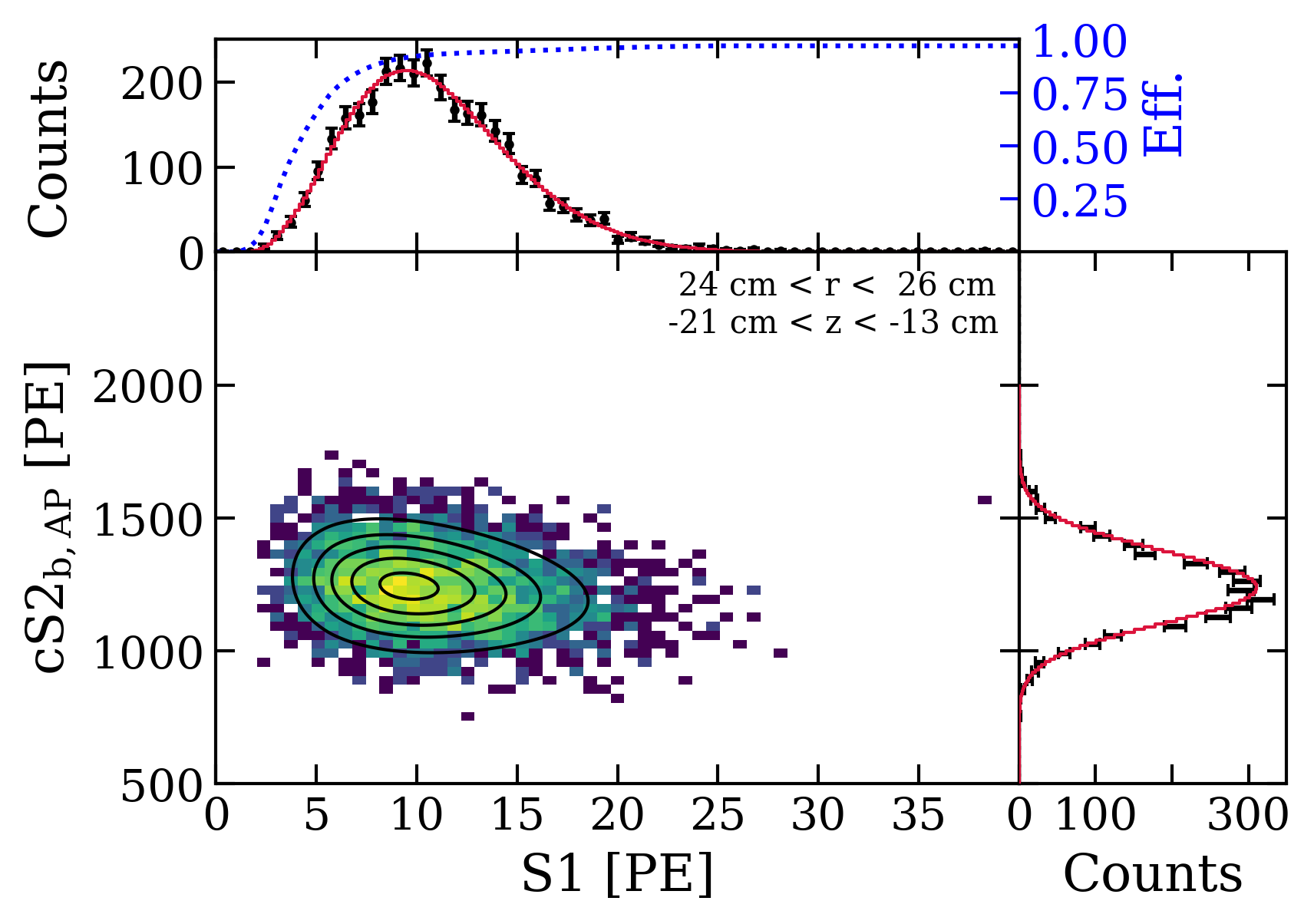}
  \includegraphics[width=0.5\textwidth]{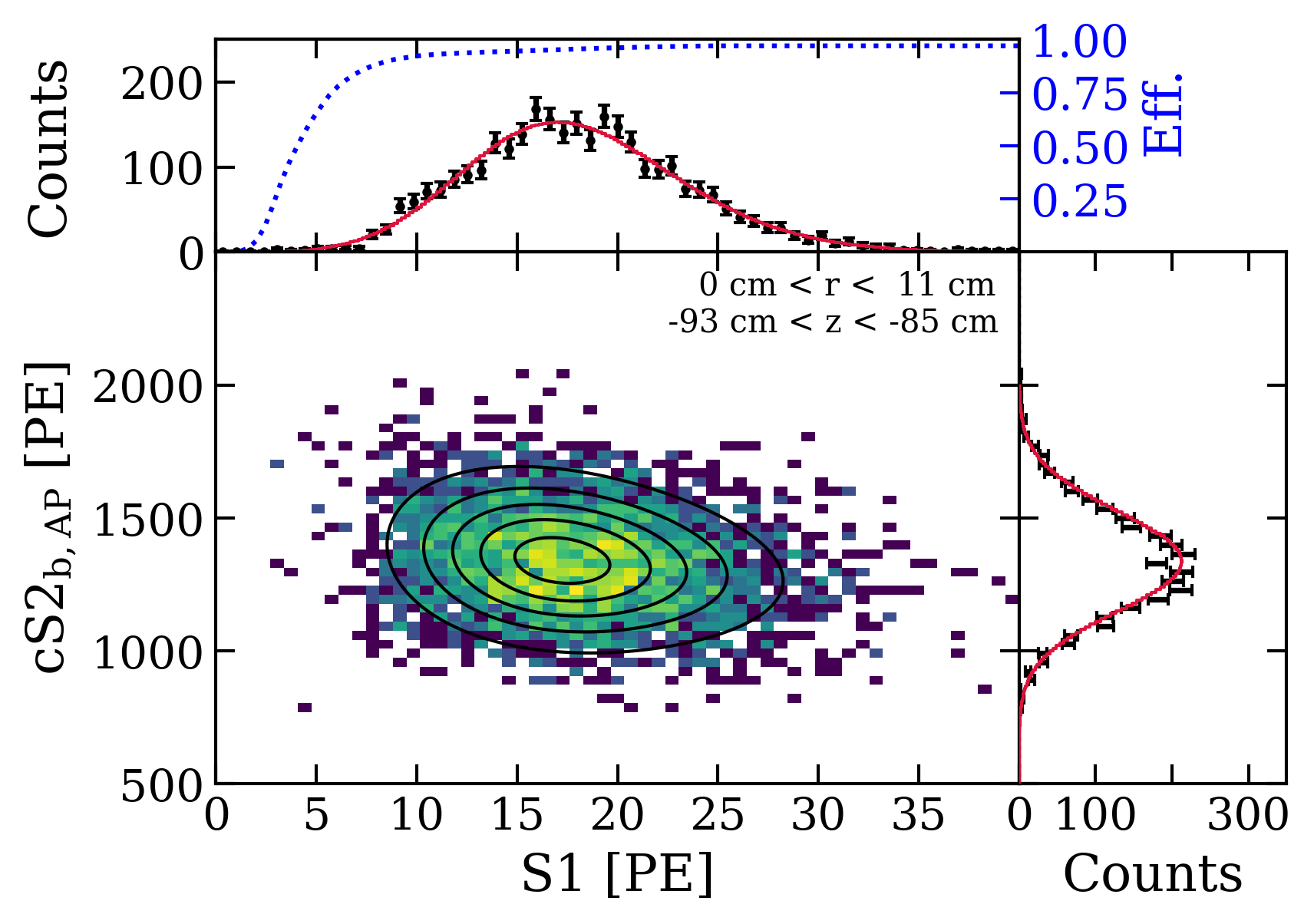}
  \caption{Direct comparison of K-shell interactions at two different positions in the TPC. The data is shown in the dataspace of drift-time and afterpulse corrected S2 v. uncorrected S1. In the top and right panels are reported the projections of the signals, with the black points for the data and the red line for the fit. The dotted blue line indicates the S1 detection efficiency. The upper plot shows events at the top and at the edge of the TPC, where the S1 detection efficiency has the highest impact. The lower plot is taken from central position at the bottom of the TPC, where the S1 is less affected by the threshold.
  %Comparison of the event distribution from the top of the TPC vs. from the bottom of the TPC in S1-cS2$_{b, AP}$ space. 
  %The S1 at the top of the TPC are much stronger affected as they are generally smaller. This is because they are mainly detected with the bottom PMT array due to reflections at the liquid surface. S2 signals detected on bottom PMT array are considered and correction includes afterpulses. In the top and right panels are reported the projections of the signals, with the black points for the data and the red line for the fit. The dotted blue line indicates the S1 detection efficiency.
  }
  \label{fig:rzfit_examples}   
\end{figure}

\label{subsec:dataAnRes}
After data selection and correction, the L$_y$ and Q$_y$ of \ar are determined via fitting in the S1-S2 space. The event distribution for a mono-energetic source is generally expected to follow a two-dimensional Gaussian distribution. However, at low energies there are non-Gaussian effects to be considered, particularly for S1 where the central limit theorem does not apply, namely:
\begin{enumerate}
    \item with $O$(10) detected photons, the Gaussian approximation to the Poisson distribution does not fully hold;
    \item the PMT single-PE response is not a Gaussian~\cite{RefModelIndependentMethod};
    \item for xenon scintillation light wavelength, PMTs \\exhibit double-PE emission, a binomial process \\with $p\sim0.2$~\cite{RefDPE, RefXE1TAnalysis}.
\end{enumerate}
Empirically, a skew Gaussian distribution was used for low-energy S1 signals, as in Ref.~\cite{RefEnRecTech,RefZeplinSkew,RefLuxSkew}.

In addition to the non-Gaussian effects described above, the sharply-dropping S1 detection efficiency for S1 $\lesssim$ 5\,PE~\cite{RefXE1TAnalysis1} is taken into account, which has a position-dependent impact on the cS1 spectrum due to spatial variations in the light collection efficiency. To include this effect, cylindrical symmetry is assumed and the LXe TPC is divided into $r$-$z$ bins of equal volumes. In each bin, a two-dimensional fit is performed in the space of uncorrected S1 and corrected S2 (cS2$_{\rm b,AP}$, b: S2 signal detected on the bottom PMT array; AP: afterpulse correction has been included). The full model, including the detection efficiency and the signal shape of the S1 distribution is given by:
\begin{linenomath*}
\begin{ceqn}
\begin{equation}
 f_1(S1) = \epsilon(S1) N e^{\frac{-(S1 - \xi)^2}{w^2}} \left(1 + \mathrm{erf}(\alpha \frac{S1 - \xi}{w\sqrt{2}})\right),
 \label{eq:s1_skew_model}   
\end{equation}
\end{ceqn}
\end{linenomath*}
where $\xi$ and $w$ are mean- and width-like parameters and $\alpha$ is related to the skewness of the distribution, with $\alpha=0$ corresponding to the standard Gaussian. The efficiency term $\epsilon(S1)$ is fixed (see Fig.~\ref{fig:rzfit_examples}) from dedicated analyses~\cite{RefXE1TAnalysis1}, and $N$ is a normalization constant. 

The cS2 spectrum is modelled using a Gaussian, and the anti-correlation between the S2 (Gaussian) mean and the S1 (skew Gaussian) mean is accounted for in the full 2D fit. Two example fits are shown in Fig.~\ref{fig:rzfit_examples}, including the S1 and cS2$_{\rm b,AP}$ projections. It can be noticed that the S1 at the top of the TPC are much more affected by the detection efficiency, as they are generally smaller. This is because they are mainly detected with the bottom PMT array due to reflections at the liquid surface. A higher value of cS2$_{\rm b,AP}$ at the bottom of the TPC is also apparent. This is because the electron lifetime from \kr, which is smaller than the one derived from \ar, is used to correct the signals. This leads to an over-correction of the \ar signals at the bottom of the TPC.

The mean of the underlying skew Gaussian in Eq.~\eqref{eq:s1_skew_model} is given by
\begin{linenomath*}
\begin{ceqn}
\begin{equation}
  \mu = \xi + \sqrt{2/\pi} \frac{\alpha w}{\sqrt{1 + \alpha^2}}, 
  \label{eq:skew_mean}
\end{equation}
\end{ceqn}
\end{linenomath*}
and is taken as the mean S1 value in each $r$-$z$ bin. The S1 means are scaled according to the light collection map derived using \kr calibration data\,\cite{RefXE1TAnalysis1} and averaged to get an overall L$_y$ to be compared with other calibration points. Q$_y$ is determined by taking the sample mean of the fitted cS2$_{\rm b,AP}$ means. In the end, the light (charge) yields are found to be
\begin{linenomath*}
\begin{ceqn}
\begin{align}
    L_y(2.82~\mathrm {keV}) &= (4.66 \pm 0.01_{\mathrm{stat.}})~\mathrm{PE/keV}, \\
    Q_y(2.82~\mathrm {keV}) &= (452.4 \pm 0.1_{\mathrm{stat.}})~\mathrm{PE/keV}.
\end{align}
\end{ceqn}
\end{linenomath*}

%To convert L$_y$ and Q$_y$ to PY and EY, $g_1$ and $g_2$ need to be determined. This is done by obtaining the L$_y$ and Q$_y$ for several other calibration points beyond \ar. Eq.~\eqref{eq:quanta_observable} can be converted to

The values of $g_1$ and $g_2$ can be obtained determining L$_y$ and Q$_y$ for other calibration points beyond \ar. Eq.~\eqref{eq:quanta_observable} can be converted to

\begin{linenomath*}
\begin{ceqn}
\begin{equation}
  Q_y = \left(-\frac{g_2}{g_1}\cdot L_y + \frac{g_2}{W}\right).
  \label{eq:doke_fit}
\end{equation}
\end{ceqn}
\end{linenomath*}

%where it becomes clear that a linear fit in the space of Q$_y$ vs L$_y$ provides $g_1$ and $g_2$.

The additional calibration points used to determine $g_1$ and $g_2$ include the injected source \kr, as well as high-energy gamma lines from material radioactivity and the activated xenon lines, $^{129\rm{m}}$Xe and $^{131\rm{m}}$Xe, which are predominately produced during neutron calibration of the detector. Since no threshold effects are applicable for these sources, the L$_y$ and Q$_y$ are determined by simple 2D (standard) Gaussian fits. 

\begin{figure}[h!]
  \centering
  \includegraphics[width=0.5\textwidth]{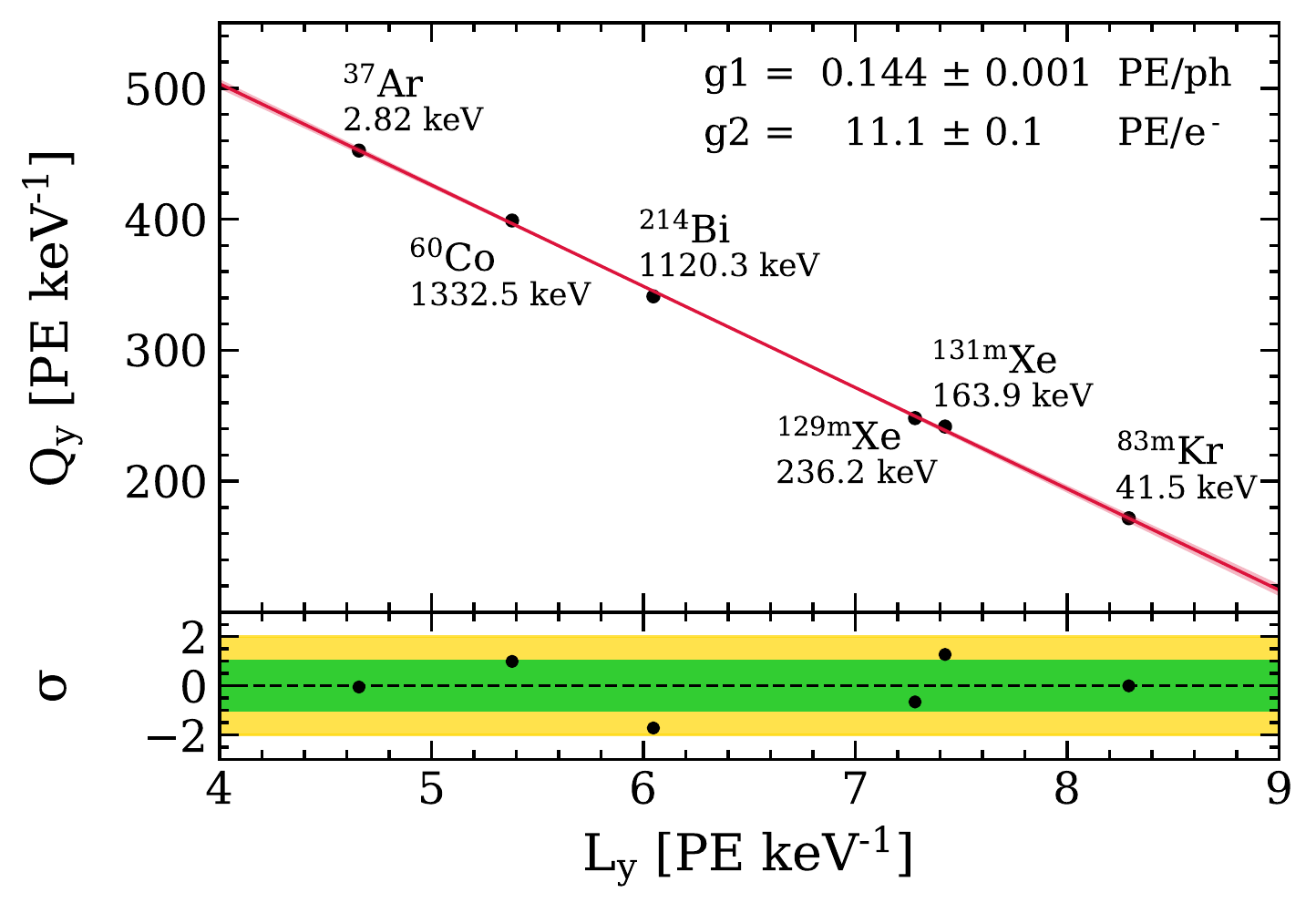}
  \caption{Charge- vs. light-yield plot with linear fit to extract the detector dependent light- and charge-gain constants $g_1$ and $g_2$. The black points represent the average S1s and S2s for various mono-energetic peaks, corrected for position-dependent effects and normalized to the energy. The red line shows the linear fit of Eq. \ref{eq:doke_fit} to the data. The red band shows the uncertainty on the fitted $g_1$ and $g_2$, dominated by systematics.}
  \label{fig:dokeFinal}   
\end{figure}

The L$_y$ and Q$_y$ values for the various calibration sources are shown in Fig.~\ref{fig:dokeFinal}, where it can be seen that a linear energy response is observed over a wide energy range, including the low-energy \ar point.

Due to the high statistics for each calibration point, the uncertainty is dominated by systematics. In order to evaluate them, the value of $\chi^2$ is set equal to the number of degrees of freedom. Assuming that every calibration points is affected by the same systematics, they are estimated to be around 0.6\%, as the values that satisfy this condition. Then a simple Monte-Carlo simulation is used to propagate these uncertainties on $g_1$ and $g_2$. $L_y$ and $Q_y$, sampled many times from Gaussian distributions which reflect such systematics, are fitted to evaluate the distribution of $g_1$ and $g_2$, whose standard deviation is taken as uncertainty.
\iffalse The systematic uncertainty was estimated by increasing the error bars in Fig.~\ref{fig:dokeFinal} until the value of $\chi^2$ equals the number of degrees of freedom and then using Monte Carlo methods to propagate the uncertainty on $g_1$ and $g_2$.\fi Potential systematic effects include small time dependencies not fully accounted for as well as the small drift field inhomogeneity hypothesized to explain the discrepancies in electron lifetime between different sources.

The linear fit to extract $g_1$ and $g_2$, taking into account these systematic uncertainties, yields
\begin{linenomath*}
\begin{ceqn}
\begin{align}
    g_1 &= (0.144 \pm 0.001)~\mathrm{PE/ph}, \\
    g_2 &= (11.1 \pm 0.1)~\mathrm{PE/e}^{-}.
\end{align}
\end{ceqn}
\end{linenomath*}

With $g_1$ and $g_2$ determined, the detector independent photon and electron yields at the \ar K-shell energy  are  
\begin{linenomath*}
\begin{ceqn}
\begin{align}
\mathrm{PY}(2.82~\mathrm{keV}) &= (32.3 \pm 0.3)~\mathrm{ph/keV}, \\
\mathrm{EY}(2.82~\mathrm{keV}) &= (40.6 \pm 0.5)~\mathrm{e}^{-}\mathrm{/keV}. 
\label{eq:EYS1S2}
\end{align}
\end{ceqn}
\end{linenomath*}

These values are shown in Fig.~\ref{fig:ArLyCy} and Fig.~\ref{fig:ArYieldField} at the end of the paper, together with expected values from NEST (Noble Element Simulation Technique) \cite{RefNestv2.3.6} and the results from other measurements at different energies and drift fields. The photon yield is in disagreement ($>5\upsigma$) with the NEST prediction at the same drift field, while it is in agreement within 2$\upsigma$ with measurements taken at same \cite{RefXurich37Ar} or similar energies \cite{RefLuxTritium}. A $>3\upsigma$ tension is also apparent with the values reported by PIXeY \cite{RefPixey}, which is not explained by the weak dependence on the drift field (Fig. \ref{fig:ArYieldField}, left panel). 
The measured electron yield is consistent with the predictions and other measurements at similar energies within 2$\upsigma$.

\begin{figure}[h!]
  \centering
  \includegraphics[width=0.5\textwidth]{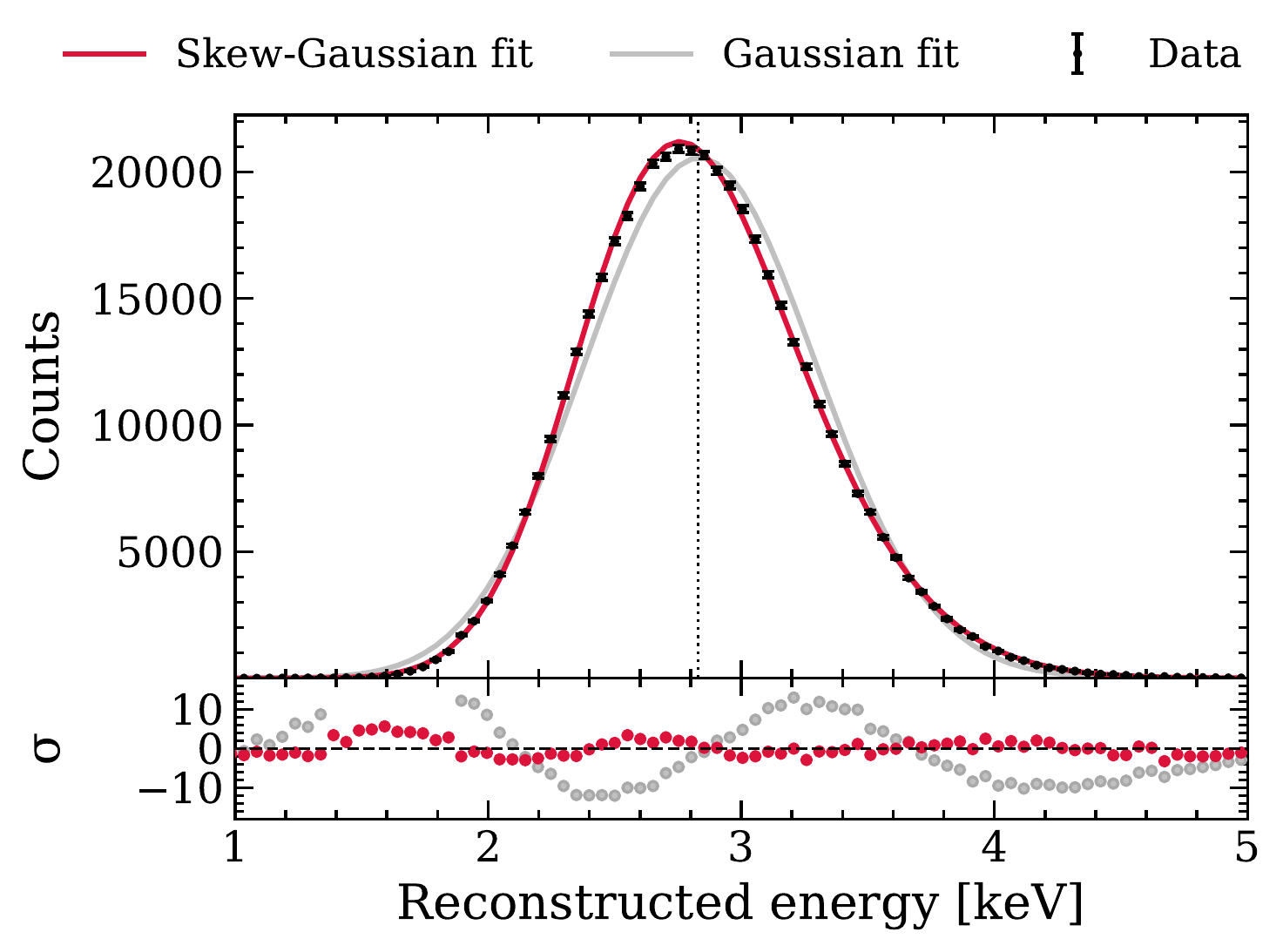}
  \caption{Reconstructed energy spectrum of \ar K-shell events using both cS1 and cS2 signals. S2 correction include also afterpulses. The black points show the data, the gray line is a Gaussian fit, and the red line is a skew Gaussian fit, as described in the text. The dotted vertical line shows the mean expected energy of the K-shell transition of 2.82\,keV.}
  \label{fig:ar37_ces}   
\end{figure}

Lastly, with the $g_1$ and $g_2$ values found above, the energy of any event can be estimated using cS1 and cS2 signals. The reconstructed energy spectrum of \ar K-shell captures is shown in Fig.~\ref{fig:ar37_ces}, along with 2 separate fits: a standard Gaussian and a skew Gaussian. As expected, the standard Gaussian does not describe the data as well as the skew Gaussian for this low-energy source. The mean of the skew Gaussian is ($2.83 \pm 0.02$) keV, consistent with the literature value $2.82$ keV and with the uncertainty dominated by the systematics of $g_1$ and $g_2$. The full-width half-maximum of the K-shell peak is $(1.04 \pm 0.01)$ keV.

\section{Implications for the XENON1T ER excess}
\label{sec:enrec}
The analysis reported here is relevant with respect to a previous result of XENON1T in which an excess of low-energy ER events was reported~\cite{RefXE1TExcessER}.
As mentioned in Sec.~\ref{sec:intro}, XENONnT data has shown no ER excess \cite{RefXEnTER}. However, it is still useful to reconsider the XENON1T ER data in the context of the improved knowledge obtained from the \ar calibration. The running conditions of the detector at the time of \ar calibration were mostly the same as for the scientific run in which the excess was observed. In particular drift field, temperature, pressure, charge yield and light yield were stable. This makes it possible to directly compare the data from the two periods. 

The model used to fit the data is updated and compared to the one applied to the results reported in Ref. ~\cite{RefXE1TExcessER}. %, in order to further validate the XENON1T results. 
The improvements included in the model are (a) the effect of PMT afterpulsing in low-energy S2s and (b) the non-Gaussian response of the detector at low energies. 
A third effect is the impact of field inhomogeneity leading to different electron lifetime values for different sources, as mentioned in Sec.~\ref{sec:datacorr}. However, the robust energy reconstruction described in Sec.~\ref{subsec:dataAnRes} suggests that this effect does not have a large impact on mean reconstructed energy. Thus, its impact is implicitly included in the empirical resolution model. %It likely does affect the energy resolution, but that impact is implicitly included in the empirical resolution model already adopted.

\subsection{PMT afterpulsing}
%As discussed in Sec.~\ref{sec:s1s2}, an $O$(1) percent negative bias is observed in \ar K-shell S2s in the top $\sim$ 20 cm of the detector. This bias was caused by the lack of PMT afterpulses reconstructed with the primary S2 for these low-energy events; for the bottom 80 cm of the LXe TPC and for higher energy events (e.g. \kr), the afterpulses were reconstructed with the primary S2.
The impact of the afterpulse correction (see Sec.~\ref{sec:datacorr}) in the top 20 cm of the detector was investigated to cross-check the energy reconstruction as reported in Ref.~\cite{RefXE1TExcessER}. The PMT afterpulsing rate was monitored throughout the operation of XENON1T. There were PMTs that displayed large increases in afterpulses consistent with xenon ions, suggesting these PMTs had small leaks that degraded their performance over time. The dataset analyzed in Ref.~\cite{RefXE1TExcessER}, called SR1, was recorded from February 2017 to February 2018, while the \ar calibration was performed in October 2018. Therefore, any bias from PMT afterpulsing is expected to be smaller in SR1 than in the calibration reported in this work.

%SR1 data are reconsidered by applying the AP correction.
The afterpulse correction impacted 26\% of events in the SR1 dataset with reconstructed energy below 15 keV, almost entirely in the top half of the LXe TPC, as expected. The S2 size for these events was increased on average by 3.3\%, with a maximum change of 6.4\%. This corresponds to the size of the deficit shown in the residuals of Fig. \ref{fig:elife}. 

\subsection{Non-Gaussian detector response}
In the SR1 analysis~\cite{RefXE1TExcessER} a Gaussian response was assumed over the full energy range for simplicity. However, as discussed in Ref.~\cite{RefEnRecTech} and shown in this work, a skew Gaussian is a more realistic model at low energies. As this change impacts mono-energetic spectra most, an improved peak search is performed in the SR1 data with a new empirical resolution model. 

The new empirical model is built using the calibration points of \ar, \kr, $^{131\rm m}$Xe, and $^{129\rm m}$Xe. A skew Gaussian model (Eq. \eqref{eq:s1_skew_model}) is used for all peaks instead of the typical Gaussian, and the skewness parameter ($\alpha$) and width parameter ($w$) are then fit as a function of the energy. The energy dependence of the skewness is modelled as a power law, while the width is modelled in the same way as the standard deviation of a Gaussian in similar contexts~\cite{RefXE1TExcessER}.  Those empirical fits are given by

\begin{linenomath*}
\begin{ceqn}
\begin{equation}
    \begin{aligned}
    \alpha(E) / E &= 2.41(1) \cdot E ^{-1.30(1)},
    \end{aligned}
\end{equation}
\end{ceqn}

\begin{ceqn}
\begin{equation}
    \begin{aligned}
     w(E) / E  &= \frac{0.374(1)}{\sqrt{E}} + 0.005(1).
    \end{aligned}
\end{equation}
\end{ceqn}
\end{linenomath*}

with $E$ expressed in keV.

\begin{figure*}[h!]
    \centering
    \includegraphics[width=1.0\textwidth]{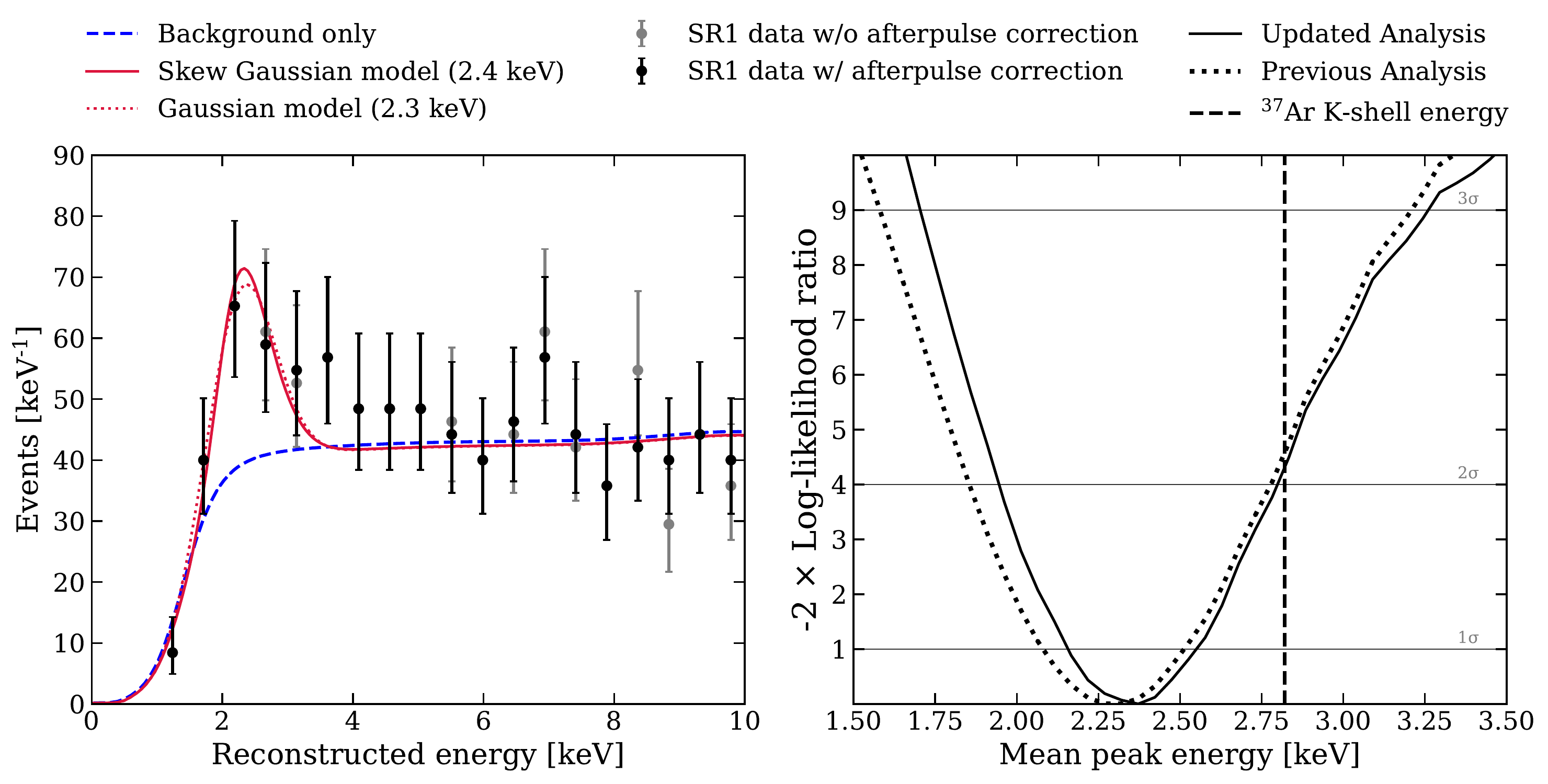}
    \caption{Updated search for mono-energetic peaks in the XENON1T SR1 data. \textit{Left panel}: SR1 energy spectrum with energy reconstructed both with the afterpulse correction (black data points) and without afterpulse correction (gray data points). The red solid line shows the best fit to the black data points for a mono-energetic peak with the updated skew Gaussian model. The red dotted line shows the best fit to the gray data points for a mono-energetic peak with the previously-used Gaussian model, as done in \cite{RefXE1TExcessER}. The dashed blue line shows the background-only fit to the black data points. \textit{Right panel}: Log-likelihood ratio for different mono-energetic peaks for both the updated analysis using the skew Gaussian and including the afterpulse correction (solid black) and the previous analysis with a standard Gaussian without afterpulse correction (dotted gray). }
    \label{fig:updated_search}
\end{figure*}

\subsection{Updated fit to SR1 ER data}
% \sloppy
%With the changes described above in reconstruction and modelling, the peak search in XENON1T background data is reconsidered here. 
A simplified version of the unbinned likelihood ratio analysis in Ref.~\cite{RefXE1TExcessER} is performed, using the same exposure, background models, and search region, but without the partitioning of the dataset in time. %The energy reconstruction is modified by including the afterpulse correction, which, as shown in Fig.~\ref{fig:updated_search} (left), has a small impact. The signal model was updated according to the new empirical smearing model using a skew Gaussian instead of a Gaussian. 
%\fussy

The result of the updated search for a mono-energetic peak is shown in Fig.~\ref{fig:updated_search}. The best-fit mean is $(2.4 \pm 0.2)$ keV, to be compared with $(2.3 \pm 0.2)$ keV from the original analysis~\cite{RefXE1TExcessER}. This indicates a small impact of the updated analysis on the energy reconstruction. The likelihood ratio curve for the peak mean is shown in Fig.~\ref{fig:updated_search} (right). The \ar peak is disfavoured at $\sim2\sigma$ with the updated analysis, as it was already found with the previous analysis~\cite{RefXE1TExcessER}.
%The updated analysis suggests relatively worse compatibility of the data with peaks $\lesssim 2$ keV on account of the skew Gaussian modelling, and it suggests that an \ar background cannot describe the excess, in addition to its exclusion via production or leakage~\cite{RefXE1TExcessER}.
In conclusion, these changes have no impact on the interpretation of the ER data in XENON1T. 

\section{S2-only analysis}
\label{sec:s2only}
\begin{figure}[h!]
  \centering
  \includegraphics[width=0.5\textwidth]{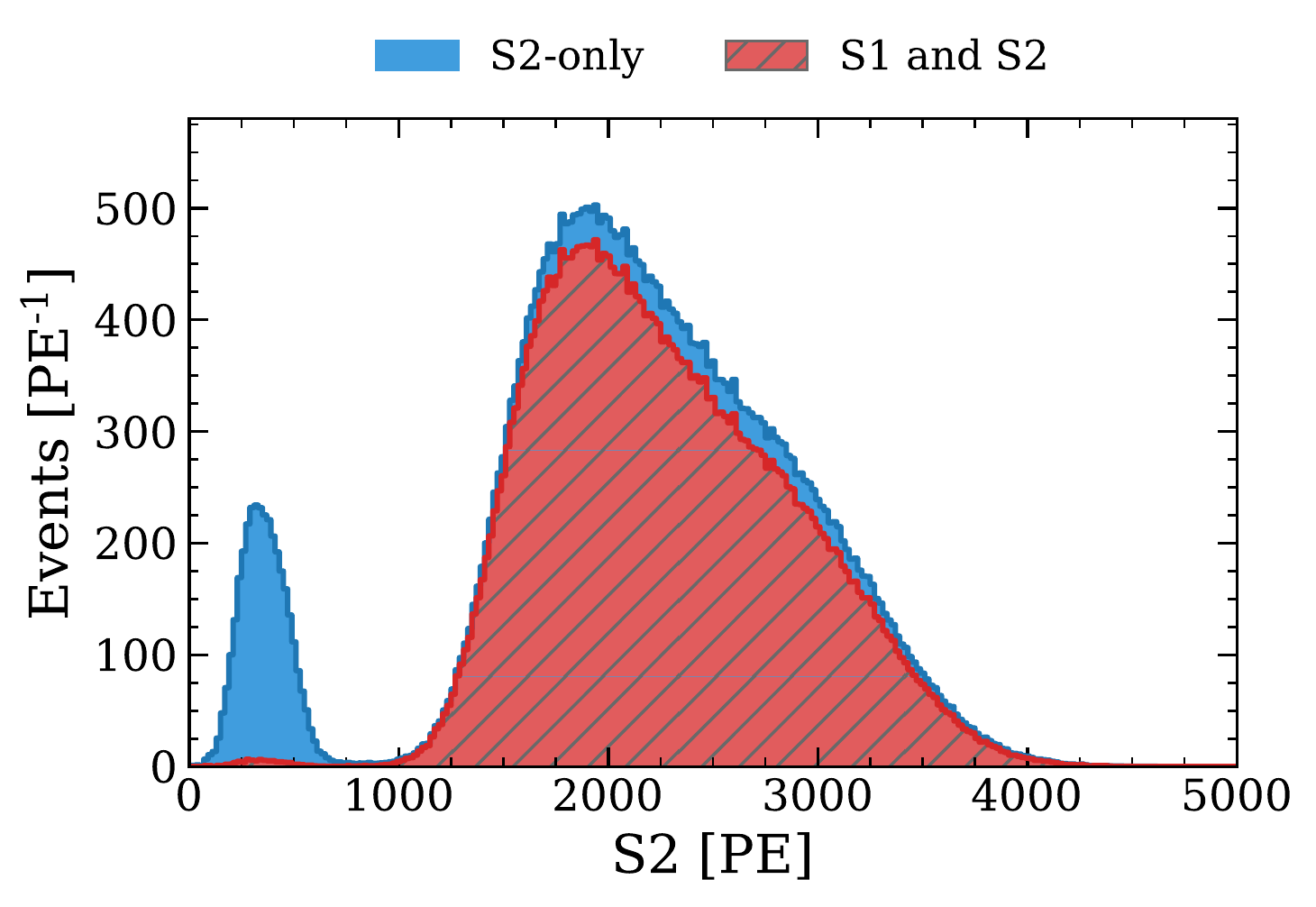}
  \caption{The \ar uncorrected S2 spectrum with the standard requirement of both S1 and S2 (red, hatched) and dropping the S1 requirement (blue). }
  \label{fig:s2only_motivation}

\end{figure}

Any S1s from the \ar L-shell decay (0.27 keV) are expected to fall well below the XENON1T energy threshold, which is determined by the 3-fold PMTs coincidence. However, the S2s  are still detectable on account of the electron amplification in the gas phase. For this reason, the study of L-shell decay can only be performed dropping the requirement of having an S1, with an S2-only analysis. The XENON collaboration already made use of this technique to search for light dark matter signals in XENON10, XENON100 and XENON1T \cite{RefXE10S2only,RefXE100S2only, RefXE1TS2only, RefXE1TSingleElectron}. Fig. \ref{fig:s2only_motivation} shows the impact of dropping the S1 requirement, which results in a L-shell peak with S2 below 1000 PE. The asymmetric shape of the peaks is due to electron absorption during drift which here is not corrected. It results that the fraction of K-shell events lost due to the S1 detection efficiency is about 8\%. The events visible at the L-shell position with S1 and S2 are the results of a random S1 (e.g. dark counts) which is accidentally paired with a S2 from L-shell decay.

While this approach allows a substantial improvement of the energy threshold, it implies a worse energy resolution and no measurement of the depth of the interaction using the drift time information. This in turn results in poorer fiducialization and hence weaker background rejection. 

The S2-only analysis is strictly necessary to study the L-shell, but it is also applied to the K-shell region as a cross check with the results from the standard S1-S2 analysis described in Sec. \ref{sec:s1s2}.\\

\subsection{Data selection}

The S2-only analysis is performed on a 6.9 days dataset acquired in the stable period after closing the krypton source and before distilling (day 7 to 14 after first injection). %A different period for this analysis with respect to S1-S2 analysis has been chosen in order to avoid contamination from \kr events.

The total S2 signal recorded by the top and bottom PMT-arrays is considered in this analysis, differently from the S1-S2 analysis. It is not possible to correct the S2 size depending on the interaction depth $z$. The S2s are only corrected depending on their $x$-$y$ positions to account for position-dependent light collection efficiency. Moreover, to be consistent with the S1-S2 analysis, the afterpulse correction is applied, see Sec. \ref{sec:datacorr}. 
Nevertheless, the relationship between S2 width and depth for interactions at a given energy can be estimated with simulations. In this way, even for events without S1, a $z$ coordinate can be statistically assigned depending on the signal width even in the L-shell energy region.

The pre-selections of data already listed in Sec. \ref{sec:s1s2DataSel} are applied and events in which an S2 is recorded are selected. Events with S2 size lower than 90 PE ($\sim$ 3 extracted electrons) are excluded.
Events greater than 5000 PE, where \ar signals are not expected, are also not used in this analysis. 

The quality cuts applied are similar to those used in the search for light dark matter \cite{RefXE1TS2only, RefXE1TSingleElectron},  but they are adapted to the specific characteristics of \ar signals, as described in the following: 

\begin{itemize}
\item[--]{\it Radial}. Events reconstructed at a radius of $r\,>\,26.5\,\,$cm are removed. \iffalse This cut is tighter than the corresponding cut in the S1-S2 analysis, due to the worse position reconstruction. 
Indeed, since $z$ is missing, it is impossible to evaluate precisely the position where the interaction occurs and, hence, the exact value of the drift field. Consequently, the field distortion correction to the position cannot be applied. Additional,  where field non-uniformities are larger (at higher radii), events tend to be mis-reconstructed.
Therefore, it is necessary to select an inner region of the TPC to be safe against these mis-reconstructed inward events, in addition to events happening close to TPC boundaries.
\fi
This cut is tighter than the corresponding cut in the S1-S2 analysis. Since $z$ is not reconstructed, field distortion corrections cannot be applied and, hence, an inner region of the TPC where the field is more uniform must be selected.

\item[--]{\it S2 Width}. \iffalse Since the width of the S2 waveform is correlated with $z$ due to diffusion of the electrons during drift, events with an S2 width outside [400, 3000] ns are removed. This cut excludes events with atypically narrow S2s, which result from radioactive decays on the electrodes at the top of the TPC, or events from the cathode region with too large widths. \fi
Events with an S2 width outside [400, 3000] ns are removed. Since the width of S2 waveform is correlated with $z$, due to electrons diffusion during drift, excluded events correspond to events at the top and the bottom of the TPC.

\item[--]{\it S2 Area Fraction Top}. On average $\sim$64\% of S2 light is expected to be seen by the top PMT array, given the detector geometry and the position where signals are produced. A selection was applied to remove events where this fraction is $>$69\%, which are due to interactions in the gaseous xenon phase above the usual secondary scintillation region (`gas events').

\item[--]{\it Pile-up}. Two dedicated cuts are implemented against pile-up of randomly emitted single-electron, which can be misidentified as S2s from real events. First, events whose S2 hit-pattern on the top array indicate two or more spatially separated small S2s which are reconstructed as a single event, as determined by a likelihood test, are removed. Second, events with two or more S2s or single electrons signals up to $\sim$1 ms before the largest S2 are not taken into account. This additionally removes cluttered waveforms, where reconstruction becomes difficult.

\item[--]{\it S2 Single Scatter}.  A selection is applied to remove multiple scatter events, in which the secondary S2 has a hit-pattern compatible with that of a scatter following a primary interaction .

\item[--]{\it Cathode}. For events where multiple S1s are present, it is checked if one of them paired with a main S2 indicates a reconstructed $z$ at the cathode position. In this case the events are excluded.
\end{itemize}

\begin{figure}[h!]
  \centering
  \includegraphics[width=0.5\textwidth]{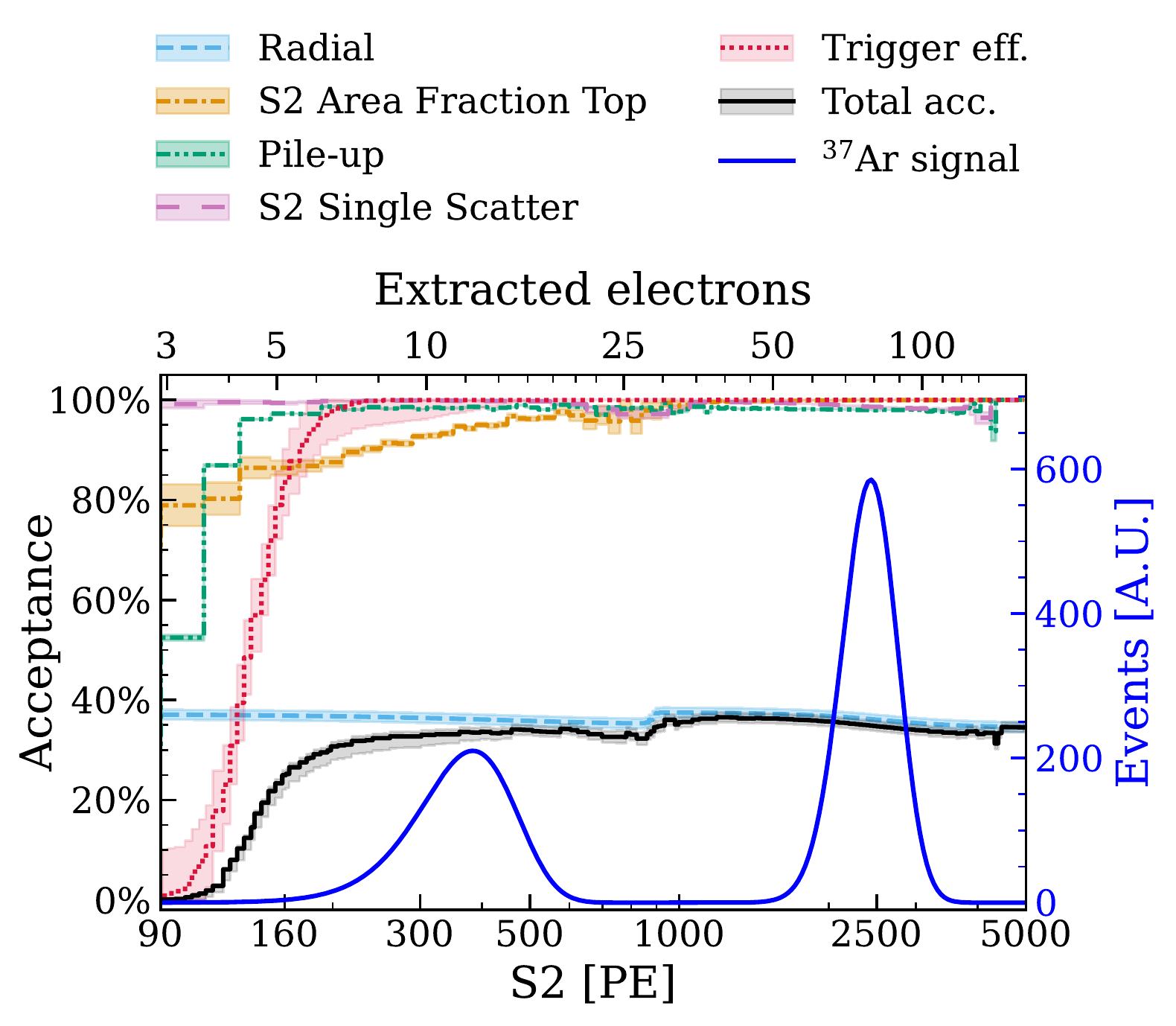}
  \caption{Acceptances (fraction of \ar events accepted) of the quality cuts and trigger efficiency depending on S2 size for the slice $-50 \;\text{cm}\,<\,z\,<\,-30\,$ cm. The bands around the lines represent the $\pm\,1\,\sigma\,$ uncertainties. The \ar signal model (without acceptances applied) is shown with a blue solid line. The number of extracted electrons corresponding to S2 is reported on top axis.}
  \label{fig:cut_acc}   
\end{figure}

%The acceptances of these cuts are calculated in the S2-$z$ space using calibration data (when possible) and the signal model (described in detail later). 
The acceptances of the quality cuts and the trigger efficiency in a given $z$ slice ($-50\,<\,z\,<\,-30\,$ cm) are shown in Fig. \ref{fig:cut_acc}, together with the signal model. {\it S2 Width} and {\it Cathode} cuts have a $\sim$ 100\% acceptance for the complete S2 range in this $z$ slice, and they are not included in this plot. The {\it Radial} cut features the lowest acceptance, but it is almost flat with respect to S2 size. 
Due to the field distortion, events at the bottom of the TPC are reconstructed more inward compared to events at the top, and are less affected by {\it Radial} cut. This produce the step around 1000 PE in the transition region, where L-shell events at the top and K-shell events at the bottom of the TPC overlap.

The trigger efficiency, namely the ability to detect S2s depending on their size, has the largest impact in the lowest S2 region.
The background contribution after the quality cuts are applied is negligible in the whole S2 size range. 

\subsection{Signal modeling}

The expected signal from \ar is evaluated following the same approach as in Ref.~\cite{RefXE1TS2only}. Since the S2 signal is uncorrected, it is not possible to use a simple Gaussian as it is done for the S1-S2 analysis. Instead, a parametric model is considered which describes the response of the detector to the \ar energy deposition in the (S2, $z$) space. 
This model is built in three steps.

The first step is the calculation of the rate of events $r_{p}(n_e)$ with $n_e$ produced electrons:

\begin{linenomath*}
\begin{ceqn}
\begin{equation}
\label{eq:s2only_model1}
r_p(n_e) = \sum_{i=K,L}\,\,(\,R_i \times \,  \mathrm{Binom}(n_e \,|\, N=(E_i/W), p_i)\,) \, ,
\end{equation}
\end{ceqn}
\end{linenomath*}

where $R_i$ is the rate of K-shell ($i=K$) or L-shell ($i=L$) decays, $N$ the number of quanta produced, $E_i$ the energy of the decay, $W$ the average energy per quanta and $p_i$ is the probability of producing an electron, with $p_i\,=\,EY_{i}\,\cdot\,W$,  where $EY_i$ is the electron yield at the K-shell or L-shell energy.% with 73 representing the total number of produced quanta per keV (photons + electrons), given the $W$ value adopted in this work.

The second step is the calculation of the rate $r_d$ of events with $k$ extracted electrons, which depends on $z$:

\begin{linenomath*}
\begin{ceqn}
\begin{equation}
\label{eq:s2only_model2}
r_d(k, z) = \sum_n \; r_p(n_e) \cdot \mathrm{Binom}\left(k \,|\, n_e, p = \eta e^{-\frac{z }{v_d \tau}} \right) \, ,
\end{equation}
\end{ceqn}
\end{linenomath*}

where $\eta$ is the extraction efficiency, $v_{d}$ the electron drift velocity and $\tau$ is the electron lifetime.

The third step is the calculation of the final rate of events $R(S2,z)$:
\begin{ceqn}
\begin{linenomath*}
\begin{equation}
\label{eq:s2only_model3}
\begin{split}
R(S2, z) 
& = \sum_k  \; r_d(k, z) \cdot \mathrm{Gauss}(S2 \,|\, \mu(k)\, , \sigma(k)) \, ,
\end{split}
\end{equation}
\end{linenomath*}
\end{ceqn}

where $\mu(k)\,=\,b(k)\cdot(\mu_{\mathrm{SE}}\cdot k)$, $\mu_{\mathrm{SE}}$ is the single electron gain (i.e. PE detected per extracted electron) and $b(k)$ is a reconstruction bias estimated from simulations, while 
\begin{linenomath*}
\begin{ceqn}
\begin{equation}
\sigma(k)\,=\,  \sqrt{\left(\,k\cdot\sigma^{2}_{\mathrm{SE}}\,+\,(\mu\cdot s_{b}(k))^{2}\,+\,(c\cdot\mu)^{2}\right)}    \, ,
\end{equation}
\end{ceqn}
\end{linenomath*}

where $\sqrt{k}\cdot \sigma_{\mathrm{SE}}$ is the resolution for $k$ electrons, $s_{b}(k)$ is the gaussian smearing of $b(k)$ and $c \cdot \mu$ is the smearing accounting for $x$-$y$ dependence of S2 ($c\,=\,0.05$ as estimated from calibrations). 

The parameters of the model have been set to the best-fit values from SR1 analysis in XENON1T~\cite{RefXE1TAnalysis} with the exception of the single electron gain $\mu_{\mathrm{SE}}\,=\,30.9\,~\mathrm{PE/e}^{-}$ which has been calculated from the value of $g_2$ reported in Sec. \ref{subsec:dataAnRes}. Five parameters, namely the two rates ($R_K$, $R_L$), the two electron yields ($EY_K$, $EY_L$) and the electron lifetime (for an independent estimate), are left free to vary and are determined from the fit to the data.

The cut acceptances and trigger efficiency are applied to $R(S2, z)$ and then the predicted signal is projected onto uncorrected S2 to compare with the data.

\subsection{Results}

The model is fit to the data in the uncorrected S2 space with a binned maximum likelihood method. The fit is restricted to events whose depth, estimated on the basis of the model from S2 size and width , corresponds approximately to $-85\, \text{ cm} \,<\,z\,<\,-12\,$ cm, in order to exclude the top and bottom part of the LXe TPC. Moreover, it is performed on the central 95\% region of the two peaks, since the model cannot perfectly describe the outer tails due to \iffalse In fact, it does not include position-dependent effects on S2 in the ($x$-$y$) plane, such as \fi non accurate corrections of the field effects which also depend on $z$.

An extensive check of the events outside the central 95\% fit region is performed. The temporal evolution of these events and their uniform spatial distribution inside the TPC are the same as for events in the central regions. It is found that these events originate from \ar decays but are poorly reconstructed in $x$-$y$ and thus end up in the signal tail region. The background contribution in the signal tails is also estimated to be negligible.

\begin{figure}[h!]
  \centering
  \includegraphics[width=0.5\textwidth]{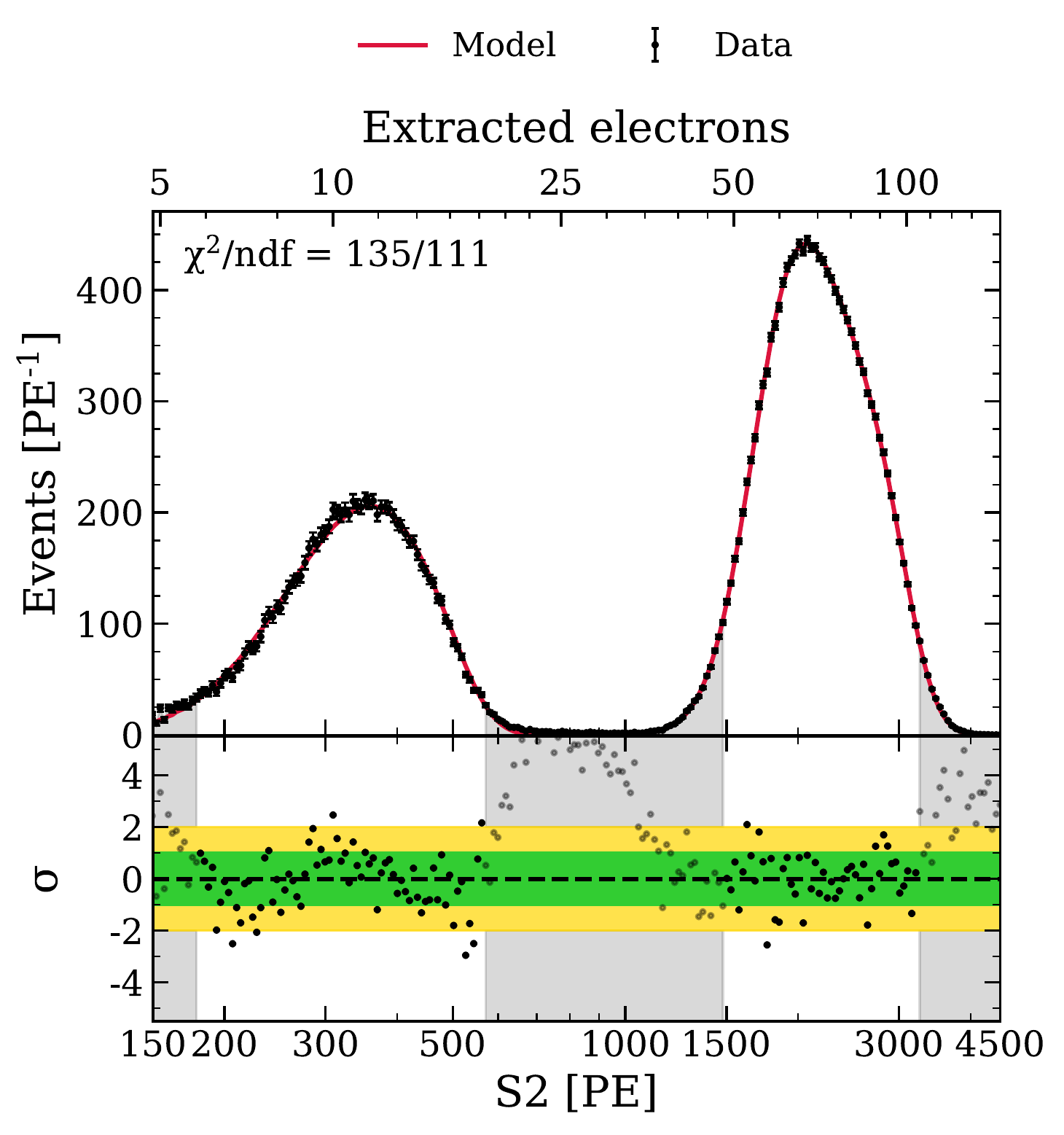}
  \caption{Best-fit model (red line) of S2 spectrum compared with data (black points). On top axis the number of extracted electrons corresponding to S2 size is reported. The gray-shaded regions are excluded from the fit, which is performed in the central 95\% region of the two peaks.%The bottom panel shows the residuals, expressed in units of standard deviation for each bin, between data and best-fit model.
  }
  \label{fig:fit_data}   
\end{figure}

The comparison of the best fit with the data is shown in Fig. \ref{fig:fit_data}. The agreement is good for both L-shell and K-shell peaks ($\chi ^2 / \text{ndf} = 135/111$). %The slight deviations on the L-shell peak are due to the resulting value of the electron lifetime, which is mainly driven by the K-shell distribution and, hence, non optimal for the L-shell region, at whose energy we expect less impact from field non-uniformity. 
There is a slight mismatch for the L-shell peak. This is a result of the electron lifetime, which is mainly driven by the K-shell distribution and, hence, non-optimal for the L-shell region, at whose energy less impact from field non-uniformity is expected. This value of the electron lifetime agrees with that based on \ar  obtained with the standard procedure (see \ref{sec:datacorr}), which only makes use of events from the K-shell.

%($\sim 930 \,\mu$s)

Systematic uncertainties on the values obtained from the fit are studied by considering the variation of the parameters of the model inside their range of uncertainty~\cite{RefXE1TAnalysis}. The main contributor to the systematic uncertainties for the electron yields is the value of the single electron gain $\mu_{\mathrm{SE}}$, which has an uncertainty of 6\% given by the sum in quadrature of the uncertainty on $g_{2}$ and of the reconstruction bias. This occurs as a result of the anti-correlation between EY and the single electron gain, for a given S2.
Varying other parameters in the model has a minor impact on the systematic uncertainties on the electron yield and on the branching ratio between electron captures of the L- and K-shell (L/K ratio). This results in a $\sim$7\% systematic uncertainty on the electron yields and a $\sim$4\% systematic uncertainty on the L/K ratio.

The electron yield best fit values for the \ar peaks are:
\begin{linenomath*}
\begin{ceqn}
\begin{align}
\mathrm{EY}(0.27~\mathrm{keV}) &= 68.0^{+6.3}_{-3.7}~\mathrm{e}^-  /\, \mathrm{keV} \, ,
\end{align}
\end{ceqn}
\begin{ceqn}
\begin{align}
\mathrm{EY}(2.82~\mathrm{keV}) &= 40.0^{+2.7}_{-2.2}~\mathrm{e}^- /\, \mathrm{keV} \, .
\end{align}
\end{ceqn}
\end{linenomath*}

\begin{figure*}[h!]
  \centering
  \includegraphics[width=0.85\textwidth]{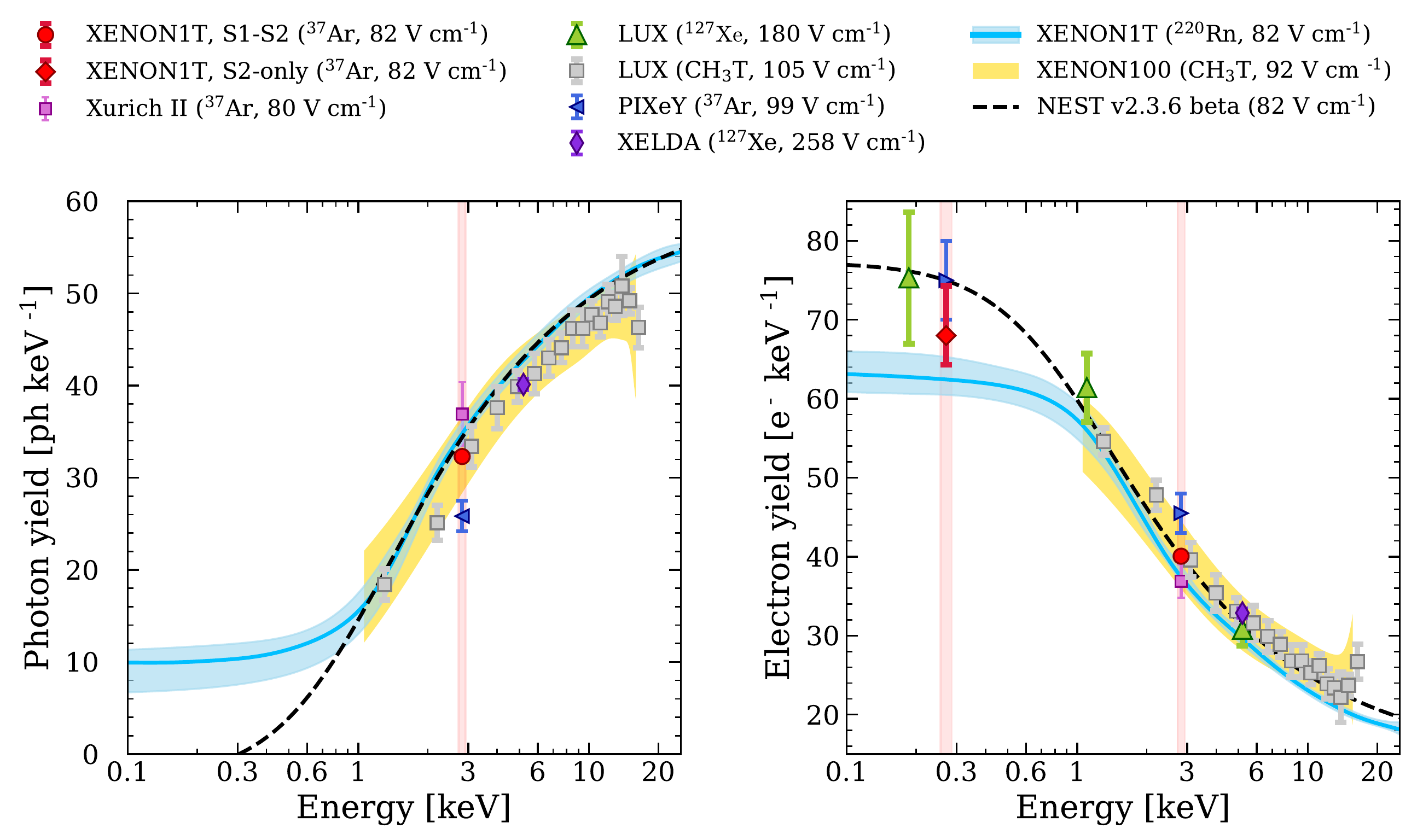}
  \caption{Photon yield (\emph{left panel}) and electron yield (\emph{right panel}) depending on energy for ER. The results obtained in this work for \ar are reported as red points, with vertical red bands highlighting the energy region corresponding to \ar decays. Are also reported the expected values from NEST v2.3.6 \cite{RefNestv2.3.6}  (black dashed line) and measurements from LUX $^{127}$Xe \cite{RefLux127Xe} and CH$_3$T \cite{RefLuxTritium}, XELDA  $^{127}$Xe \cite{RefXelda127Xe} , PIXeY  \cite{RefPixey} and Xurich II \cite{RefXurich37Ar} $^{37}$Ar, XENON100 CH$_3$T \cite{RefXenon100Tritium}, as well as the fit to the XENON1T calibration data from $^{220}$Rn (light blue solid line, shaded region represents the 15\%-85\% credible region) \cite{RefXE1TAnalysis}.}
  \label{fig:ArLyCy}   
\end{figure*}

\begin{figure*}[h!]
  \centering
  \includegraphics[width=0.85\textwidth]{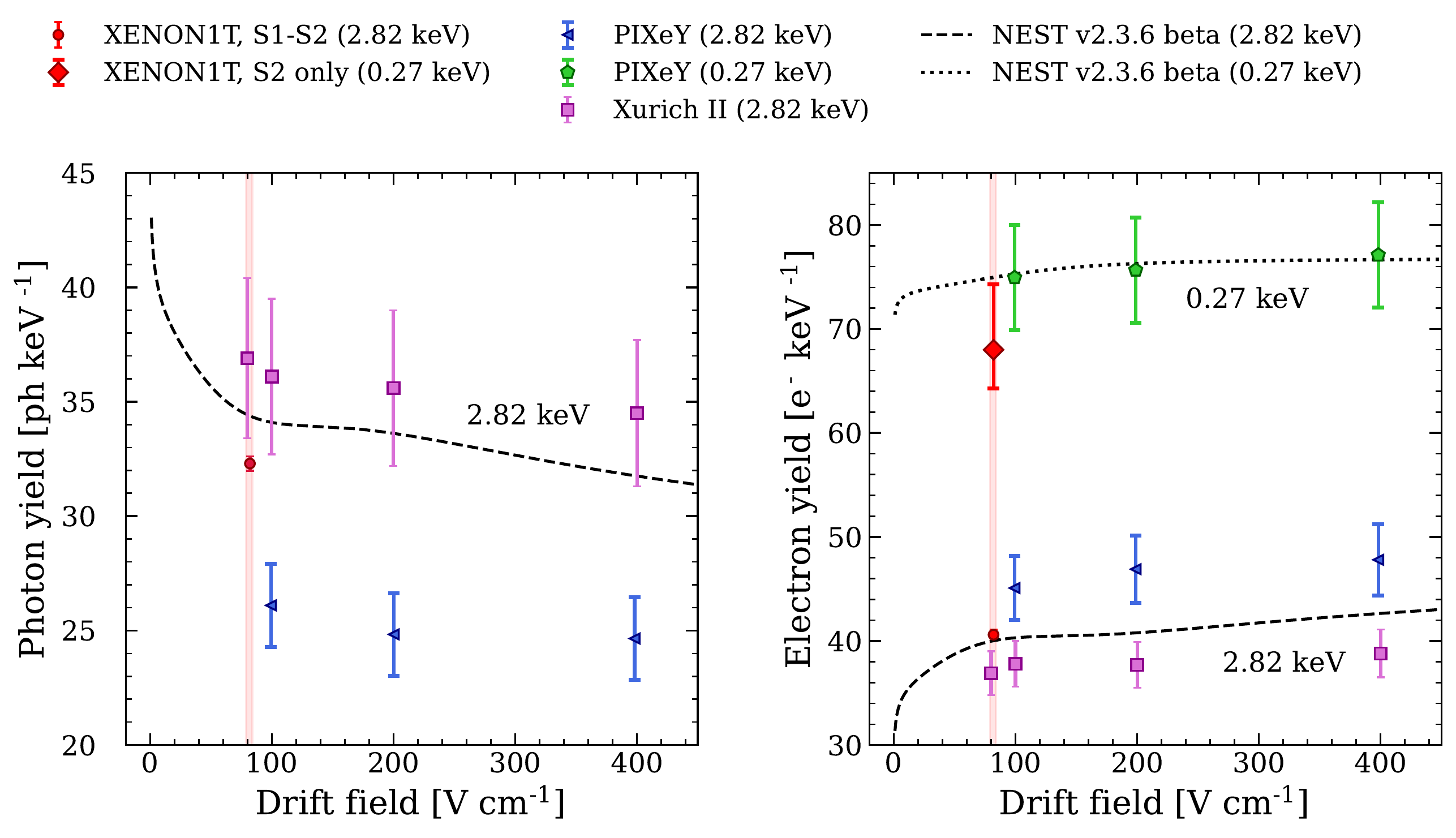}
  \caption{Photon yield (\emph{left panel}) and electron yield (\emph{right panel}) at different drift fields for \ar peaks obtained by PIXeY \cite{RefPixey} and Xurich II \cite{RefXurich37Ar} compared to the results from this work (red points). The vertical red bands indicate XENON1T drift field. The blue triangles indicate yield at 2.82 keV while the green pentagons represent electron yield at 0.27 keV. The expected values from NEST v2.3.6 \cite{RefNestv2.3.6} are shown with a black dashed line (2.82 keV) and a black dotted line (0.27 keV). 
}
  \label{fig:ArYieldField}   
\end{figure*}

 The electron yield at 2.82 keV is in agreement with the result from S1-S2 analysis, see Eq. \eqref{eq:EYS1S2}. As shown in Fig. \ref{fig:ArLyCy} and Fig. \ref{fig:ArYieldField}, the electron yield at 0.27 keV is compatible with both NEST predictions \cite{RefNestv2.3.6} and with the extrapolations from the $^{220}$Rn calibration in XENON1T \cite{RefXE1TAnalysis}. It is also in agreement within 1$\upsigma$ with the measurement by PIXeY at a slightly higher drift field (99 V/cm) \cite{RefPixey}. It adds information in the low-energy region ($<\,1$ keV), where only few measurements are available \cite{RefPixey,RefLux127Xe}. It has to be noted that no subtraction of the electrons produced directly by the decay and then drifted to the gas phase has been applied, differently from what performed in Ref. \cite{ds50}. This allows a direct comparison with all the other measurements reported in this work, in which such a correction is also not applied.

The L/K ratio obtained from the best-fit model is

\begin{linenomath*}
\begin{ceqn}
\begin{equation}
\mathrm{L/K}\, = \, (10.11 \pm 0.44) \%   \, 
\end{equation}
\end{ceqn}
\end{linenomath*}

 in agreement within 1$\upsigma$ with the expected value of 9.67\%~\cite{RefTable37Ar, RefAr37energy} \iffalse evaluated from the shells' transition probability\fi. This suggests that the S2 detection efficiency and cut acceptances in the energy range between 0.27 keV and 2.82 keV are well understood.
Furthermore, the good description of the \ar signal for both peaks validates the modelling used for S2-only analysis in XENON1T \cite{RefXE1TS2only}.

\section{Conclusions}
% condensed version, don't know if its better, feel free to adapt/rewrite
\label{sec:conclusions}
A calibration was performed with the low-energy lines of \ar at 2.82 keV (K-shell decay) and 0.27 keV (L-shell decay)  in XENON1T. The mandatory removal of this source after calibration via cryogenic distillation was successfully demonstrated, reaching a reduction of activity by one order of magnitude every 4.5 days. This proves that \ar can be used as a regular calibration source for multi-tonne xenon detectors.

Two analyses of the \ar data are performed, one using both S1 and S2 signals, the other only requiring the detection of the S2. At the XENON1T drift field of 82 V/cm, the photon and electron yields at 2.82 keV are $(32.3 \pm 0.3)\,$photons/keV and $(40.6 \pm 0.5)\,$electrons/keV, respectively. The electron yield at 0.27 keV is $68.0^{+6.3}_{-3.7}\,$electrons/keV.

\iffalse
\begin{linenomath*}
\begin{ceqn}
\begin{align}
\mathrm{PY}(2.82~\mathrm{keV}) &= (32.3 \pm 0.3)~\mathrm{photons/keV} , \\
\mathrm{EY}(2.82~\mathrm{keV}) &= (40.6 \pm 0.5)~\mathrm{electrons/keV} , 
\end{align}
\end{ceqn}
\end{linenomath*}
and the electron yield at 0.27 keV is
\begin{linenomath*}
\begin{ceqn}
\begin{align}
\mathrm{EY}(0.27~\mathrm{keV}) &= 68.0^{+6.3}_{-3.7}~\mathrm{electrons/keV} .
\end{align}
\end{ceqn}
\end{linenomath*}
\fi
Since only a weak dependence from the drift field is expected at these energies, these values can be compared with measurements taken at different drift fields. As it is apparent from Fig. \ref{fig:ArLyCy} and Fig. \ref{fig:ArYieldField}, there is in general a good agreement with other measurements at the same or similar energies. In particular agreement within 1$\upsigma$ is observed among XENON1T and LUX, the two large detectors providing measurements at these energies. Some tension is observed with the measurements from PIXeY at 2.82 keV, which is much larger than the difference due to the higher drift field, whereas yields from Xurich II are closer to the results of this work. Disagreement is also apparent for the photon yield at 2.82 keV with respect to the NEST prediction at the same drift field. The electron yield value at 0.27 keV is in agreement with predictions and other measurements at the same or similar energies. This new measurement adds relevant information in an energy region where only two other experimental values are currently available. 

The energy of the K-shell decay reconstructed with the combined energy scale is (2.83$\pm$0.02) keV, in agreement with the expected value. The results from the \ar calibration suggests that XENON1T detector is well understood even in the low-energy ER region close to the detection threshold.

The outcomes of this work are used to inform a re-analysis of XENON1T science data in which a low-energy ER excess was previously reported. With the  additional correction to the data and the updated resolution model described in this work, the main conclusions obtained in Ref. \cite{RefXE1TExcessER} remain unaltered.

\section*{Acknowledgements}

We gratefully acknowledge support from the National Science Foundation, Swiss National Science Foundation, German Ministry for Education and Research, Max Planck Gesellschaft, Deutsche Forschungsgemeinschaft, Helmholtz Association, Dutch Research Council (NWO), Weizmann Institute of Science, Israeli Science Foundation, Binational Science Foundation, Fundacao para a Ciencia e a Tecnologia, R\'egion des Pays de la Loire, Knut and Alice Wallenberg Foundation, Kavli Foundation, JSPS Kakenhi and JST FOREST Program in Japan, Tsinghua University Initiative Scientific Research Program and Istituto Nazionale di Fisica Nucleare. This project has received funding/support from the European Union’s Horizon 2020 research and innovation programme under the Marie Sk\l{}odowska-Curie grant agreement No 860881-HIDDeN. Data processing is performed using infrastructures from the Open Science Grid, the European Grid Initiative and the Dutch national e-infrastructure with the support of SURF Cooperative. 
We are grateful to Laboratori Nazionali del Gran Sasso for hosting and supporting the XENON project.

%\begin{figure}
%  \includegraphics{example.eps}
%\end{figure}
%\begin{figure*}
%  \includegraphics[width=0.75\textwidth]{example.eps}
%\end{figure*}
%\begin{table}
%\begin{tabular}{lll}
%\hline\noalign{\smallskip}
%\end{tabular}
%\end{table}

%\begin{acknowledgements}
%\end{acknowledgements}



\end{document}